\documentclass[useAMS, usenatbib]{mnras}
\usepackage{graphicx,amsmath,color,amssymb}

\usepackage[pdftitle={}]{hyperref}
\usepackage[dvipsnames]{xcolor}

\usepackage{fontawesome}

\definecolor{flatirons}{HTML}{8B2131}
\definecolor{sunshine}{HTML}{CA9500}
\definecolor{skyline}{HTML}{1D428A}
\definecolor{midnight}{HTML}{0E2240}

\newcommand{\beq}{\begin{equation}}
\newcommand{\eeq}{\end{equation}}
\newcommand{\barr}{\begin{eqnarray}}
\newcommand{\earr}{\end{eqnarray}}

\newcommand{\Lya}{Lyman-$\alpha$}

\newcommand{\mfemu}{\texttt{MFEmulator}}
\newcommand{\mfbox}{\texttt{MF-Box}}

\newcommand{\Data}{\mathcal{D}}

\newcommand{\gp}{\textsc{gp}}

\newcommand{\GP}{\mathcal{GP}}

\newcommand{\normal}{\mathcal{N}}

\newcommand{\uniform}{\mathcal{U}}

\newcommand{\mpgadget}{\textsc{mp-gadget}}

\newcommand{\yvec}{\boldsymbol{y}}

\newcommand{\thetavec}{\boldsymbol{\theta}}

\newcommand{\Kvec}{\boldsymbol{\mathrm{K}}}
\newcommand{\kvec}{\boldsymbol{k}}

\newcommand{\Mpch}{\,\textrm{Mpc\,}h^{-1}}
\newcommand{\hMpc}{\,h\textrm{Mpc}{^{-1}}}

\newcommand{\lowres}{LF} 
\newcommand{\highres}{HF} 

\newcommand{\npart}{N_\mathrm{p}}

\newcommand{\xyemulator}[2]{#1\,{\lowres}-#2\,{\highres} emulator}

\newcommand{\outputFunction}{f}
\newcommand{\outputVector}{\yvec}

\begin{document}

\title[MF-Box: Multi-Fidelity with varying box sizes]{
MF-Box: Multi-fidelity and multi-scale emulation for the matter power spectrum
}
\author[ M.-F. Ho et al.]{Ming-Feng Ho,$^1$\thanks{E-mail: mho026@ucr.edu} Simeon Bird,$^1$\thanks{E-mail: sbird@ucr.edu} Martin A. Fernandez,$^1$\thanks{E-mail: mfern027@ucr.edu} Christian R. Shelton.$^2$\thanks{E-mail: cshelton@cs.ucr.edu}\\
$^1$Department of Physics \& Astronomy, University of California, Riverside,
900 University Ave., Riverside, CA 92521, USA\\
$^2$Department of Computer Science \& Engineering, University of California, Riverside, 900 University Ave., Riverside, CA 92521, USA\\
}

\date{\today}

\pagerange{\pageref{firstpage}--\pageref{lastpage}} \pubyear{2023}
\pagenumbering{arabic}
\label{firstpage}

\maketitle

\begin{abstract}
We introduce {\mfbox}, an extended version of {\mfemu}, designed as a fast surrogate for power spectra, trained using N-body simulation suites from various box sizes and particle loads. To demonstrate {\mfbox}'s effectiveness, we design simulation suites that include low-fidelity suites (L1 and L2) at $256 \Mpch$ and $100 \Mpch$, each with $128^3$ particles, and a high-fidelity suite (HF) with $512^3$ particles at $256 \Mpch$, representing a higher particle load compared to the low-fidelity suites. {\mfbox} acts as a probabilistic resolution correction function, learning most of the cosmological dependencies from L1 and L2 simulations and rectifying resolution differences with just 3 HF simulations using a Gaussian process.
{\mfbox} successfully emulates power spectra from our HF testing set with a relative error of $< 3\%$ up to $k \simeq 7 \hMpc$ at $z \in [0, 3]$, while maintaining a cost similar to our previous multi-fidelity approach, which was accurate only up to $z = 1$. The addition of an extra low-fidelity node in a smaller box significantly improves emulation accuracy for {\mfbox} at $k > 2 \hMpc$, increasing it by a factor of $10$.
We conduct an error analysis of {\mfbox} based on computational budget, providing guidance for optimizing budget allocation per fidelity node. Our proposed {\mfbox} enables future surveys to efficiently combine simulation suites of varying quality, effectively expanding the range of emulation capabilities while ensuring cost efficiency.

\end{abstract}


\section{Introduction}



Over the past decade, cosmological large-scale structure surveys have evolved increasingly in resolution and size.
As observations probe more non-linear structures with high precision,
theoretical predictions must be highly accurate to match the observational errors at corresponding small scales.
The only way to achieve such accurate predictions is by running $N$-body simulations.
However, including expensive numerical simulations in the cosmological inference will require $\sim 10^6$ likelihood evaluations using simulations, i.e., $\sim 10^6$ numerical simulations in the Markov Chain Monte Carlo (MCMC) sampling, making it impractical to use simulations for Bayesian inference directly.

In the development of statistical surrogate modeling, emulators emerged as a Bayesian approach to analyze simulations and perform fast function predictions \citep{Currin:1991,Santner:2003,OHagan:2006}.
In cosmology, emulators have been widely used as a fast surrogate model to replace the expensive likelihood evaluations in the MCMC sampling.
For example,
using surrogate models to replace the Boltzmann code in cosmological inference \citep{Auld:2007,Auld:2008,Arico:2021:Boltzmann,Alessio:2022,Nygaard:2022,Gunther:2022}. With a large number of training samples ($\sim \mathcal{O}(10^4-10^6)$), these Boltzmann code emulators have successfully improved the speed of the current parameter estimation pipeline.
Another approach is using surrogates to replace MCMC to emulate the posterior distribution directly, reducing the overall required number of likelihood evaluations \citep{ElGammal:2022}.

Unlike the emulators for Boltzmann codes, likelihood evaluations based on numerical simulations, such as cosmological $N$-body simulations, are more expensive per training sample.
Therefore, only a limited number of full-size training simulations ($\sim \mathcal{O}(10^1 - 10^2)$) are computationally available.
Emulation based on numerical simulations has been implemented in various cosmological applications:
the matter power spectrum \citep{Heitmann:2009,Heitmann:2014,Lawrence:2017,Euclid:2019,Euclid:2021},
baryonfication simulations \citep{Arico:2021,Giri:2021},
arbitrary cosmology \citep{Giblin:2019},
$f(R)$ gravity \citep{Arnold:2022,Harnois-Deraps:2022},
weak lensing \citep{Harnois:2019,Davies:2021,Giblin:2023},
halo mass function \citep{McClintock:2019,Nishimichi:2019,Bocquet:2022},
21-cm power spectrum \citep{Kern:2017} and global signal \citep{Cohen:2020,Bevins:2021,Bye:2022},
and {\Lya} forest \citep{Bird:2019,Rogers:2019,Pedersen:2020,Rogers:2021a,Rogers:2021b,Cabayol-Garcia:2023}.
All these emulators are self-consistent and can replicate the simulations as surrogate models to accelerate the parameter inference pipeline.

Emulators have also been used in several current surveys.
\cite{Zurcher:2022} used an emulator on Dark Energy Survey year 3 data (DES Y3) for cosmic shear peak statistics.
\cite{Neveux:2022} used an emulator on SDSS quasars and galaxies.
Beyond cosmological inference,
\cite{Jo:2023} uses emulation to calibrate the galaxy formation simulations.
\cite{Salcido:2023, Kugel:2023} build emulators to quantify the subgrid feedback effects in the hydrodynamical simulations.
Emulators have also been used in a wide range of disciplines,
for example, exoplanet \citep{Rogers:2021}, gravitational wave \citep{Cheung:2021},
stellar population synthesis \citep{Alsing:2020},
heavy-ion physics \citep{Ji:2021,Ji:2022},
astrochemistry \citep{Holdship:2021}, and biology \citep{Vernon:2018}.

The computational costs of cosmological emulators are rapidly increasing, driven by an increase in both survey accuracy and number of model parameters.
Over the past few years, cosmological emulators based on $N$-body simulations have evolved from five-dimensional cosmology (e.g., $w$CDM in Coyote Universe \citep{Heitmann:2009}) to higher dimensions, for example, eight-dimensional $w_0w_a$CDM$+\sum m_\nu$ cosmology in \cite{Euclid:2021} and Mira-Titan Universe \citep{Lawrence:2017,Moran:2023}.
The increase in dimensionality means the number of simulations required for training an accurate emulator also needs to increase dramatically.
For instance, EuclidEmulator2 requires more than 200 high-resolution simulations with $3000^3$ in an eight-dimensional cosmology.
Moreover, when the astrophysics effects are not ignorable for cosmological inference \citep{Giri:2021,Arico:2021,Villaescusa-Navarro:2021},
more expensive simulations, such as hydrodynamical simulations including baryonic effects, must be used for training realistic emulators.
This increase in computational cost poses a challenge for the implementation of emulators in future surveys, making them prohibitively expensive and difficult to adopt unless the efficiency of emulation techniques can be improved.


An efficient approach to reducing the computational cost is building emulators using multi-fidelity emulation ({\mfemu}), which allows simulations with different particle loads to be combined \citep{Ho:2022}.
\cite{Fernandez:2022} showed that it is possible to construct a realistic emulator using hydrodynamical simulations through the {\mfemu} technique, emulating {\Lya} forest with sub-percent test accuracy using only $6$ high-fidelity simulations.
In \cite{Ho:2022,Fernandez:2022}, we assumed the particle load is the only fidelity variable.
This is a limitation, as simulation volumes also correlate with the accuracy of a simulation: With a constant particle load, larger box sizes enhance accuracy at larger scales but diminish it at smaller scales due to reduced mass resolution. Smaller volumes with the same particle load can capture finer small-scale details, though a minimum box size requirement exists \citep{Heitmann:2010,Schneider:2016}.
Here we show that the cost of training a {\mfemu} can be further reduced by having multiple fidelities which vary both simulation volumes and particle loads.

The multi-fidelity method we use, based on \cite{Kennedy:2000,Ho:2022}, is just one of many multi-fidelity techniques.
\cite{Peherstorfer:2018} surveyed the multi-fidelity methods in uncertainty quantification, inference, and optimization.
A few popular methods include the control variate technique, which has been applied in cosmology in
\cite{Chartier:2021,Chartier:2022} on reducing the variance of the covariance matrix, and multi-level or multi-stage Markov Chain Monte Carlo \citep{Andres:2005,Lykkegaard:2020}, which use low-fidelity models to reduce the number of expensive likelihood evaluations in MCMC.
Though multi-level MCMC is a promising method, its practical use requires running thousands of $N$-body simulations in the sampler, which is not yet applicable to cosmological inference.
Another similar method is using deep learning methods to learn the mapping from low- to high-resolution simulations to directly generate the snapshots of the `super-resolution' simulations \citep{Li:2021,Doogesh:2020,Ni:2021}. 
While this method shows promise, it is currently limited to a single cosmology and is not yet suitable for inference.

The statistical and computer science literature already contains work on multi-fidelity techniques with more than one low-fidelity node. \cite{Lam:2015,MISO:2016} considered a multi-information source framework, which combines more than one information node to achieve an overall lower variance.
In this work, we use a graphical Gaussian process, based on a directed acyclic graph \citep{Ji:2021}, to predict high-fidelity simulations using low-fidelity simulations in two different simulation volumes.

A design using multiple low-fidelity nodes can be helpful in several ways.
One example, which we will show in this work, is enhancing the resolution at small scales using an additional low-fidelity node with a smaller box size.
A cosmological simulation has strict volume requirements to ensure that the base mode is linear and to beat cosmic variance.
However, it also needs high enough particle load (or spatial resolution) to capture the non-linearities at small scales.
{\mfemu} provides a way to improve small-scale structures using a simulation suite from a lower particle load.
Nevertheless, the non-linear information in a lower particle-load simulation is also limited.
An economical way to resolve small scales is to run simulations in small boxes to increase the spatial resolution by sacrificing some large-scale information.

Another approach to minimizing the number of training simulations is Bayesian optimization, where a sequential choice of new training simulations is designed to optimize the likelihood function globally.
For example, \cite{Rogers:2019,Leclercq:2018,Takhtaganov:2019} implemented Bayesian optimization in the cosmological inference.
Similar approaches, such as \cite{Pellejero:2020,Cole:2022,Boruah:2022,Neveux:2022}, iteratively train emulators on the high likelihood regions of the parameter space, thus minimizing the overall training samples to achieve accurate posterior distribution. Our multi-fidelity emulation is a complimentary technique, which can be combined with Bayesian optimisation for the lowest computational cost.

This paper presents {\mfbox}, extending our previously developed {\mfemu} to allow multiple low-fidelity nodes in a multi-fidelity emulator.
{\mfbox} uses the multi-fidelity graphical Gaussian process model (GMGP) \citep{Ji:2021} to emulate high-fidelity simulations using low-fidelity simulations from two different simulation volumes.
A GMGP model is an extension of the traditional KO model \citep{Kennedy:2000} and NARGP model \citep{Perdikaris:2017}.
The difference is that a GMGP allows multiple nodes in a fidelity while KO or NARGP models assume one node per fidelity.
For example, in our case, the low-fidelity nodes include separate box sizes with the same particle load, resolving different scales of the Universe.

Our references to low- and high-fidelity nodes are based on a relative scale within the context of our multi-fidelity framework. We do not directly compare these definitions to other matter power spectrum emulators. Our primary goal is to demonstrate the effectiveness of {\mfbox} as a probabilistic resolution correction tool. This allows us to correct the resolution of a low-fidelity emulator, approximating higher particle loads using a limited number of high-fidelity simulations.

Consequently, the focus of our discussion on emulation error revolves around predicting unseen high-fidelity simulations in the test set. This choice is intentional, as it allows us to assess how well {\mfbox} can upscale a low-fidelity emulator when predicting high-fidelity simulation outputs.
It is worth highlighting that the framework we present here can be adapted for use with various other summary statistics emulators, accommodating different definitions of low- and high-fidelity nodes as needed.

We will also present an analysis of the emulation errors in relation to the computational budget.
Previous studies \cite{Ji:2021,Wendland:2004} have demonstrated that Gaussian process emulator errors can be bounded by a power-law function.
In this paper, we model the emulation error from {\mfbox} as a power-law function of the number of training simulations and empirically infer the emulator error function from our {\mfbox} results. By utilizing this empirical error function, we can estimate the emulation error associated with a given multi-fidelity design, as well as determine the optimal budget allocation for each node.
This error analysis serves as a useful guide for future development of {\mfemu} techniques.

In Section~\ref{sec:sim}, we will describe our simulations and experimental design.
Section~\ref{sec:emulation} will review the single-fidelity emulator as well as three multi-fidelity emulation methods, namely AR1, NARGP, and {\mfbox}.
Our sampling strategy for selecting input cosmologies for high-fidelity simulations will be outlined in Section~\ref{sec:sampling}.
Empirical inference of the emulation error function will be discussed in Section~\ref{sec:budget}.
Section~\ref{sec:results} will present the results of {\mfbox}, followed by the conclusion in Section~\ref{sec:conclusions}.


\section{Simulations}
\label{sec:sim}

We perform dark matter-only simulations using the open source MP-Gadget code \citep{MPGADGET:2018},\footnote{\url{https://github.com/MP-Gadget/MP-Gadget/}}
an $N$-body and smoothed particle hydrodynamical (SPH) simulation code derived from Gadget-3 \citep{Springel:2003} and used to run the \textsc{astrid} simulation \citep{Bird:2022,Ni:2022}, a large-scale high-resolution cosmological simulation with $250 \Mpch$ containing $2 \times 5500^3$ particles.
The base of MP-Gadget is Gadget-3, but, among other improvements, it has been rewritten to take advantage of shared-memory parallelism and the hierarchical timestepping from Gadget-4 \cite{Springel:2021}.
Detailed descriptions of the simulation code can be found in \cite{Bird:2022}.

\begin{figure*}
    \includegraphics[width=2\columnwidth]{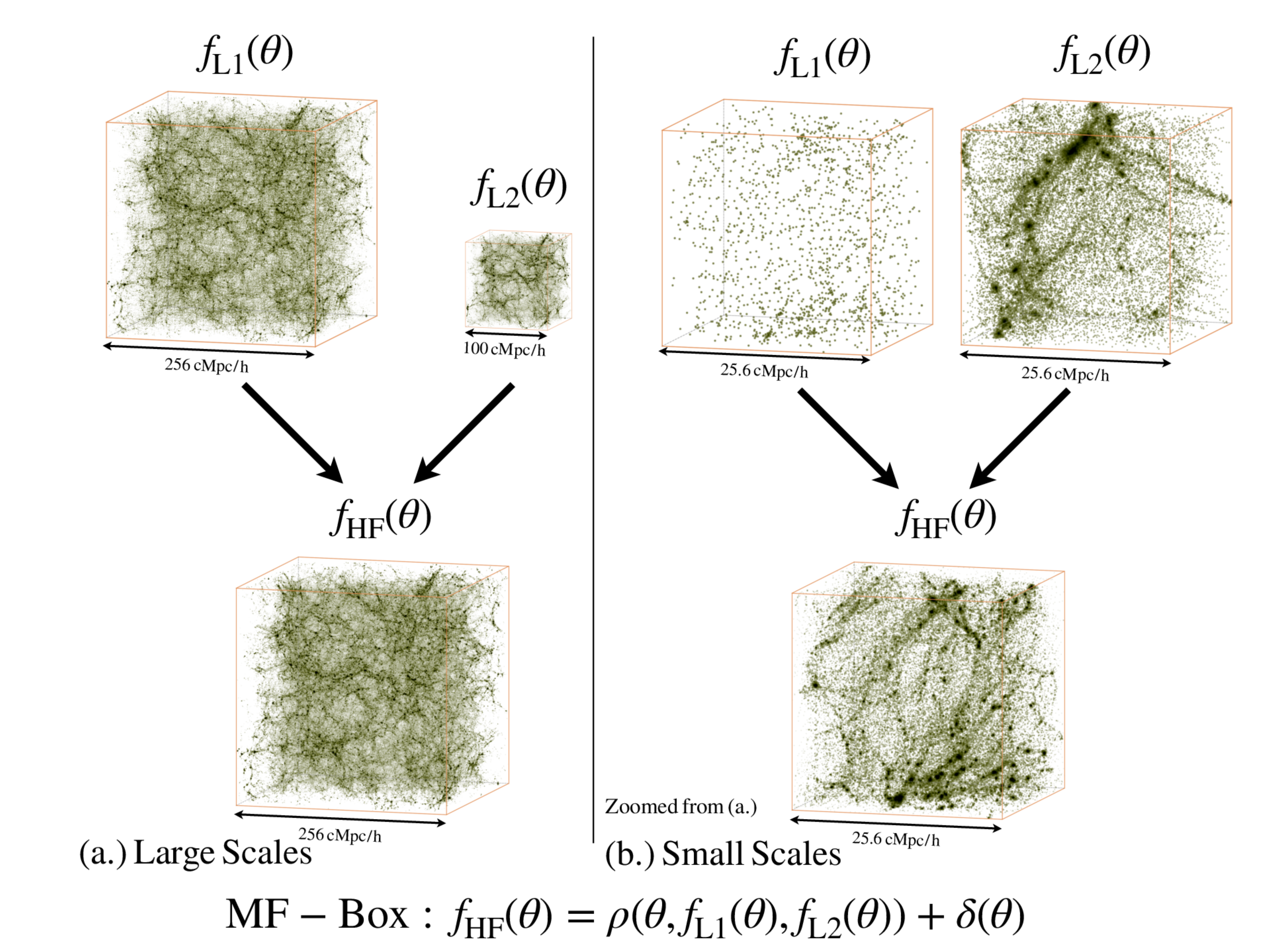}
    \caption{Illustration of the {\mfbox} framework and the dark-matter only simulations performed at $z = 0$.
    {\mfbox} provides a emulation framework to connect power spectra (denoted as $f(\theta)$, where $\theta$ is the input cosmology) from low-fidelity simulations (L1 and L2) to high-fidelity simulations (HF), providing an efficient emulation framework in predicting HF power spectra using only a few HF simulations augmented with many low-fidelity simulations with various volumes.
    $\rho$ is a learnable multiplicative resolution correction parameter, and $\delta$ is a learnable additive resolution correction parameter.
    Details of the {\mfbox} model can be found in Section~\ref{subsubsec:mfemu_emu_gmgp}.
    The particle loads and box sizes for each simulation are listed in Table~\ref{table:simulations}. \textbf{(a.)} Large-scale structures of each simulation are shown. Simulations L1 and L2 have the same particle load ($\npart = 128$), but L1 has a smaller box size ($100 \Mpch$). As a result, the large scales of L1 resemble those of the high-fidelity (HF) simulation, while L2 lacks the necessary large-scale information to match HF. \textbf{(b.)} Zoomed-in view ($25.6 \Mpch$) of the small scales from (a.). L1 lacks structures due to the sparsity of particles at this scale, whereas L2 captures more structures by utilizing a smaller box size. As a result, L1 resembles HF at small scales due to its finer mass resolution.}
    \label{fig:nbody_plot}
\end{figure*}

We start the simulations at $z = 99$ and finish at $z = 0$.
The initial linear power spectrum and transfer function are produced by CLASS \citep{Lesgourgues:2011} at $z = 99$ through the Zel'dovich approximation \citep{Zeldovish:1970}.  We assume periodic boundary conditions.
We use a Fourier-transform-based particle-mesh method on large scales for the gravitational forces and a Barnes-Hut tree \citep{BarnesHut:1986} on small scales.
Table~\ref{table:simulations} summarizes the simulation volumes and particle loads used in this paper.
We use the same set of low-fidelity (L1) and high-fidelity (HF) pairs as in \cite{Ho:2022}, with an additional low-fidelity node (L2) to demonstrate the emulation using simulations from different box sizes.
However, the framework presented in this paper is generalizable to more than two low-fidelity nodes.
Fig~\ref{fig:nbody_plot} shows a visual illustration for the dark-matter only simulations used in this paper.

\begin{table}
	\centering
	\caption{Low- and high-fidelity simulation suites used in our study. The definition of low- and high-fidelity nodes is based on a relative scale specific to our approach and is not intended for direct comparison with other matter power spectrum emulators.}
	\label{table:simulations}
	\begin{tabular}{lcccc}
		\hline
		Simulation & Box Volume & N$_{\text{part}}$ & Node Hour\\
		\hline
		L1 & $(256$ $\Mpch)^3$ & $128^3$   &  $\sim 1.0$   \\
		L2 & $(100$ $\Mpch)^3$ & $128^3$   &  $\sim 1.7$ \\
        HF & $(256$ $\Mpch)^3$ & $512^3$   &  $\sim 140$ \\
		Test & $(256$ $\Mpch)^3$ & $512^3$ &  $\sim 140$ \\
		\hline
	\end{tabular}
\end{table}

Our emulation target is the matter power spectrum, $P(k)$, a summary statistic of the over-density field.
We measure the matter power spectrum with a cloud-in-cell mass assignment.
We use the built-in power spectrum estimator from MP-Gadget; the power spectrum is thus generated on a mesh the same size as the simulation's PM grid, which is $3$ times the mean interparticle spacing.
The multi-fidelity emulation framework we introduce here is also applicable to other implementations of power spectrum calculations, such as those generated by NBodyKit \citep{Hand:2018}.

\begin{figure}
    \includegraphics[width=\columnwidth]{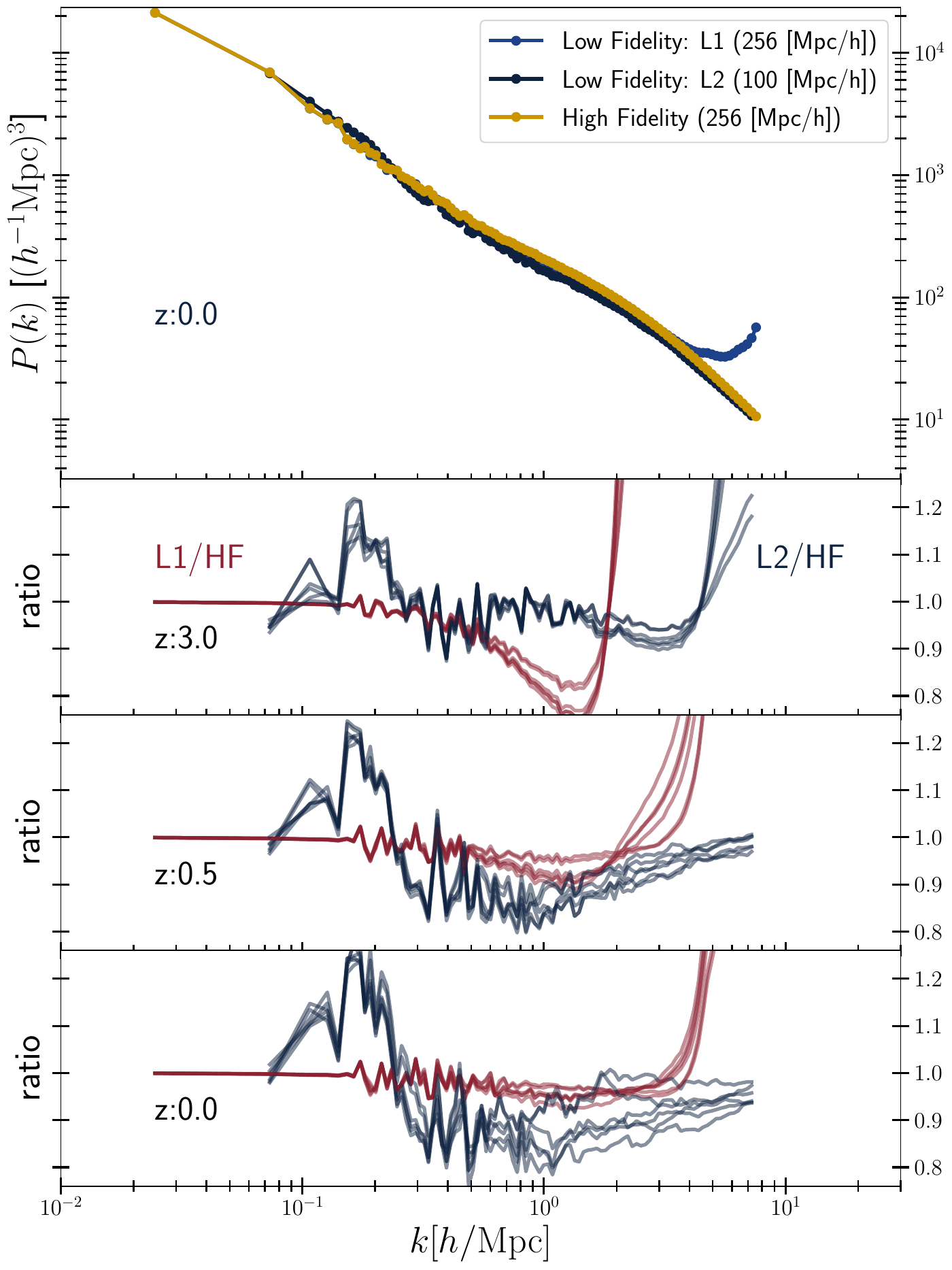}
    \includegraphics[width=\columnwidth]{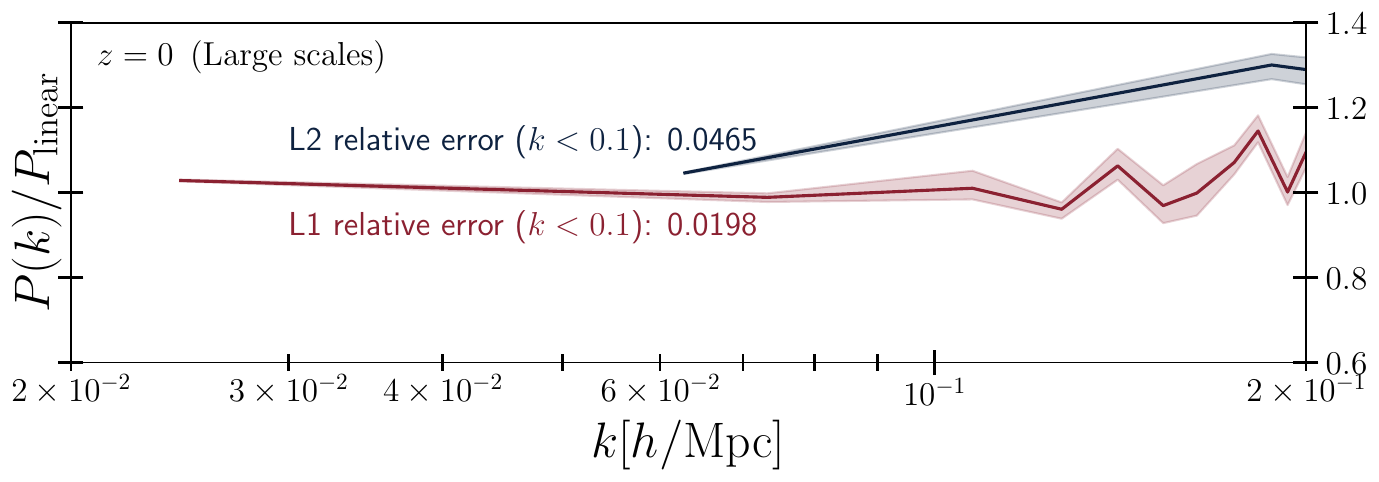}
    \caption{Matter power spectra from dark-matter only MP-Gadget simulations with various fidelities, conditioning on the same cosmology.
    The top panel shows the power spectra from a large-box low-fidelity (L1; \textcolor{skyline}{blue}), a small-box low-fidelity (L2; \textcolor{midnight}{black}), and a large-box high-fidelity simulations (HF; \textcolor{sunshine}{yellow}).
    The numeric values for different fidelities of simulations are tabulated in Table~\protect\ref{table:simulations}.
    The 2nd, 3rd, and bottom panels show the ratios of L1/HF (\textcolor{flatirons}{red}) and L2/HF (\textcolor{midnight}{black}) simulations, conditioned on different redshift bins, $z = 3.0,\, 0.5,\, 0$.
    (Bottom panel): We also show the ratio between (L1, L2) and the linear theory power spectrum from CLASS at large scales.
    The solid lines show the median and shaded areas show the $68\%$ quantiles across 60 different cosmologies.
    }
    \label{fig:powerspecs}
\end{figure}

Figure~\ref{fig:powerspecs} shows an example of our emulation target: matter power spectra from different resolutions, where the low-fidelity simulations (L1 and L2) have two different box sizes.
L1 simulations are in the same box size ($256 \Mpch$) as high-fidelity simulations (HF) with the same initial condition seeding;
whereas, L2 simulations have a smaller box size ($100 \Mpch$) than L1 and HF.
In principle, L2 can capture more small-scale structures due to its smaller box size.
Indeed, as shown in the 2nd, 3rd, and bottom panels in Figure~\ref{fig:powerspecs},
L2 is more accurate than L1 at small scales.
For example, at $z = 3$, L2/HF is closer to 1 than L1/HF at small scales ($k > 0.6 \hMpc$).

Note that L2 is not necessarily better than L1 in matching the HF simulations.
L1 matches the HF power spectrum extremely well at large scales,
while L2 performs better at small scales.
Therefore, the accuracy of the different simulations are not in a monotonically increasing sequence. Thus the \cite{Kennedy:2000} method we used in \cite{Ho:2022} cannot be directly applied to this example.

\begin{figure}
    \includegraphics[width=\columnwidth]{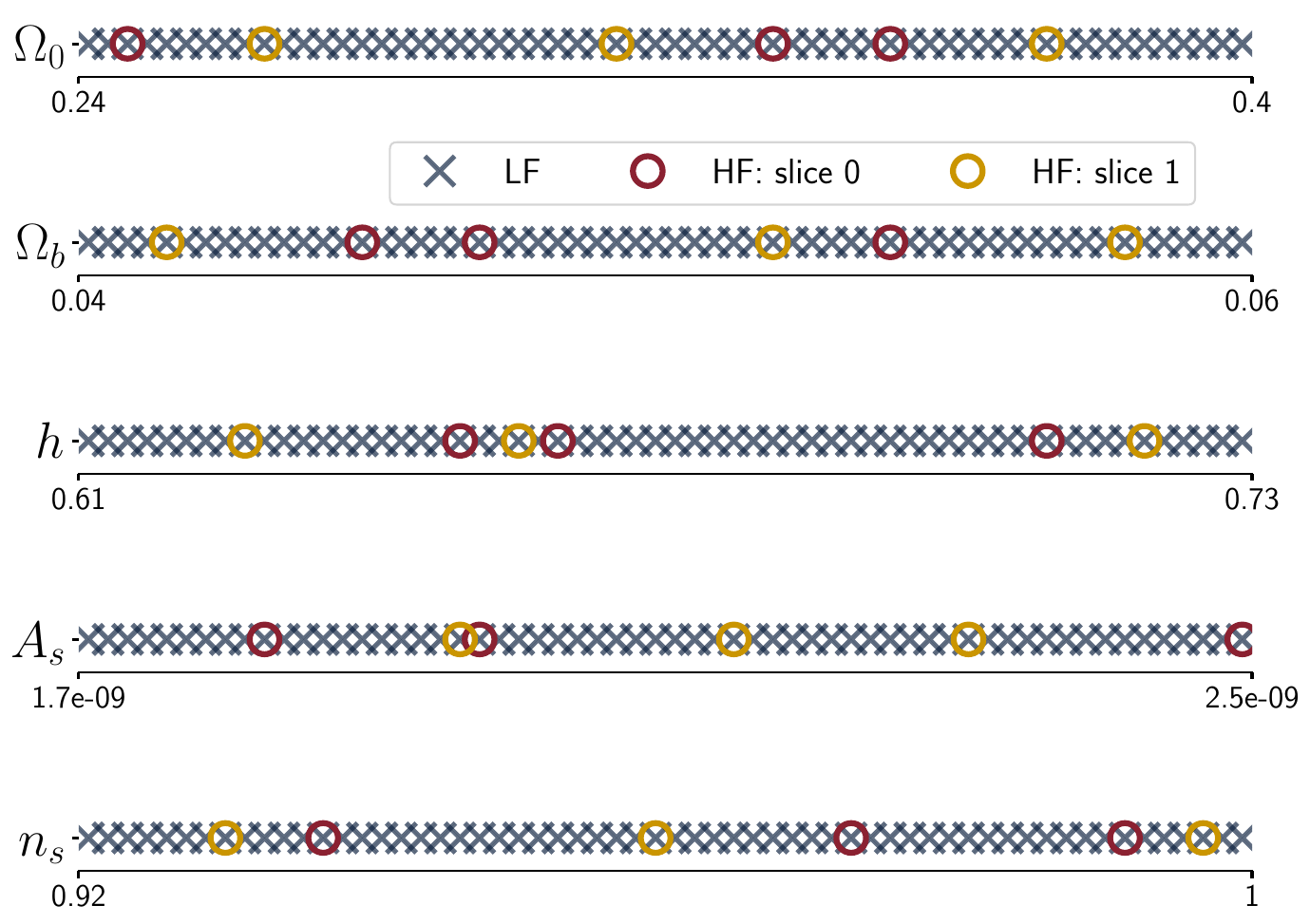}
    \caption{Experimental design of low- and high-fidelity simulations in this work.
    The prior volume is chosen to be the same as EuclidEmulator2 \protect\citep{Euclid:2021}.
    Crosses (\textcolor{midnight}{black}) are the input parameters for the low-fidelity simulations (both L1 and L2).
    Circles (\textcolor{flatirons}{red} and \textcolor{sunshine}{yellow}) are the parameters for high-fidelity simulations, which is a subset of the low-fidelity experimental design.
    We use max-min Sliced Latin Hypercube (SLHD) \citep{Ba:2015} for the LF design, containing 20 slices with 3 samples in each slice.
    \textcolor{flatirons}{Red} and \textcolor{sunshine}{Yellow} circles show two of the slices, which we select to be the input parameters for HF simulations.
    }
    \label{fig:LHS_samples}
\end{figure}

Figure~\ref{fig:LHS_samples} shows our experimental design in the input parameter space, corresponding to the prior range of
\begin{equation}
    \begin{split}
        \Omega_0 &\sim \uniform(0.24, 0.4);\\
        \Omega_b &\sim \uniform(0.04, 0.06);\\
        h &\sim \uniform(0.61, 0.73);\\
        A_s/10^{-9} &\sim \uniform(1.7, 2.5);\\
        n_s &\sim \uniform(0.92, 1),
    \end{split}
\end{equation}
where $\Omega_0$ is the total matter density parameter in the Universe, $\Omega_b$ is the total baryon density parameter, $h$ is the dimensionless Hubble parameter, $A_s$ is the spectral amplitude and $n_s$ is the spectral index.

We generated 60 Latin hypercube samples using max-min Sliced Latin Hypercube \citep{Ba:2015},
including 20 slices with 3 samples in each slice.
We will discuss SLHD in Section~\ref{subsec:slhd}.
SLHD partitions the design into several equal slices (or blocks).
Each slice itself is also a Latin hypercube design, as well as the whole design.
We thus choose one of the Latin hypercube slices as our high-fidelity input.
By using SLHD, we can avoid the design points of the HF node clustered in the corner of the prior volume.
We ran L1 and L2 nodes using the same cosmological parameters (although this is not required by the GMGP from \cite{Ji:2021}).

We summarize the notation used in this paper in Table~\ref{tab:notation}.

\begin{table}
    \caption{Notations and definitions}
    \begin{tabular}{|l|l|}
       Notation & Description \\
       \hline
       {HF} & High Fidelity \\
       {LF} & Low Fidelity \\
       $\thetavec$ & Input cosmological parameters\\
       $f(\thetavec)$ & Summary statistics (matter power spectrum\\
       &in this work) corresponding to input parameters.\\
       $\npart$ & Number of particles per box side\\
       AR1 & Autoregressive GP \\
        &\citep{Kennedy:2000}\\
       NARGP & Non-linear autoregressive GP\\
        & \citep{Perdikaris:2017}\\
       GMGP & Graphical GP \citep{Ji:2021}\\
       {\mfemu} & Multi-fidelity cosmological emulator \\
       & \cite{Ho:2022}\\
       {\mfbox} & Multi-fidelity cosmological emulator\\
       & with different box sizes in low fidelity.
    \end{tabular}
    \label{tab:notation}
 \end{table}

\section{Emulation}
\label{sec:emulation}


Emulation predicts the output from expensive cosmological simulations.
First, a handful of simulations are run at carefully chosen experimental design points as a training set.
Next, a surrogate model (an emulator) fits the prepared training set to predict simulation output.
The trained emulator will be a proxy for the simulation results, allowing for inexpensive evaluation of a likelihood function.

In Section~\ref{subsec:gp_emu}, we will briefly review emulation using a Gaussian process.
Section~\ref{subsec:mfemu_emu} will review how we can extend the Gaussian process emulator to model simulations from different qualities using a multi-fidelity emulator, {\mfemu}.
Our earlier multi-fidelity technique based on the KO method \citep{Kennedy:2000} will be reviewed in Section~\ref{subsubsec:mfemu_emu_ko}.
Section~\ref{subsubsection:nargp} will review an extension of the KO method based on a deep Gaussian process, NARGP \citep{Perdikaris:2017}.
Section~\ref{subsubsec:mfemu_emu_gmgp} describes a graphical-model Gaussian process model (GMGP) \citep{Ji:2021}, an extension of NARGP to allow more than one node in the same fidelity.

\subsection{Gaussian process emulator}
\label{subsec:gp_emu}

A Gaussian process (\gp) regression model \citep{Rasmussen05} is widely used as a cosmological emulator.
A {\gp} provides closed-form expressions for predictions. In addition, a {\gp} naturally comes with uncertainty quantification, which is handy for inference framework and Bayesian optimization.
In emulation, a {\gp} can be seen as a Bayesian prior for the simulation response.
It is a prior because the emulator model is chosen to ensure smoothness in the simulation response \textit{before} data are collected \citep{Santner:2003}.

Let $\thetavec \in \Theta \subseteq \mathbb{R}^{d}$ be the input cosmologies for the simulator,
and $\outputFunction(\thetavec)$ be the corresponding output summary statistic.
This work assumes that the summary statistic is the non-linear matter power spectrum.
A {\gp} regression model is a prior on the response surface of our simulated matter power spectrum:
\begin{equation}
    p(\outputFunction) = \GP(f; \mu, k),
\end{equation}
where $\mu(\thetavec) = \mathbb{E}[\outputFunction(\thetavec)]$ is the mean function, and $k(\thetavec, \thetavec') = \mathrm{Cov}[\outputFunction(\thetavec), \outputFunction(\thetavec')]$ is the covariance kernel function.
The mean function is usually assumed to be a constant or zero mean unless there is prior knowledge about the mean function.
In this work, we assume a zero mean function.
The covariance kernel function is typically chosen as a squared exponential function (radial basis function, RBF) to return a smooth response surface.

Suppose we run the simulations at $n$ carefully chosen input cosmologies, $\mathcal{D} = \{\thetavec_1, \cdots, \thetavec_n\}$, and we compress each simulation into the corresponding matter power spectrum, $\outputVector = \{ \outputFunction(\thetavec_1), \cdots, \outputFunction(\thetavec_n) \}$.
Conditioning on this training data and optimizing the hyperparameters using maximum likelihood estimation,
we can get the predictive distribution of $\outputFunction$ at a new input cosmology $\thetavec_*$, $\outputFunction_* = \outputFunction(\thetavec_*)$, through a closed-form expression
\begin{equation}
    p(\outputFunction_* \mid \outputVector_*, \Data, \thetavec)
    = \normal(\outputFunction_* \mid \mu_*(\thetavec_*), \sigma_*^2(\thetavec_*)),
\end{equation}
where the mean and variance are
\begin{equation}
    \begin{split}
        \mu_*(\thetavec_*)
        &= \kvec(\thetavec_*, \Data)^\intercal \Kvec(\Data)^{-1} \outputVector;\\
        \sigma_*^2(\thetavec_*) &= k(\thetavec_*, \thetavec_*) - \kvec(\thetavec_*, \Data)^\intercal \Kvec(\Data)^{-1} \kvec(\thetavec_*, \Data).
    \end{split}
\end{equation}

The vector $\kvec(\thetavec_*, \Data) = [ k(\thetavec_*, \thetavec_1), \cdots, k(\thetavec_*, \thetavec_n) ]$ represents the covariance between the new input cosmology, $\thetavec$, and the training data.
The matrix $\Kvec(\Data)$ is the covariance of the training data.

Although we do not explicitly state this in the notation, we let $\outputFunction(\thetavec)$ be a single-value output. If the target summary statistic is a vector, we let the Gaussian process model each bin separately. It will be more apparent why we make this modeling decision in later sections (Section~\ref{subsec:mfemu_emu}). The primary reason is that the correlation between low-fidelity and high-fidelity summary statistics changes depending on the scales. The multi-fidelity method can only capture scale dependence if we model the scales separately.\footnote{An alternative way is to apply a co-kriging kernel to model the covariance for each vector element. We do not do that in this work because we found the single-output {\gp} is enough for our cosmological emulation purpose, so there is no need to introduce another layer of complexity.}

\subsection{Multi-Fidelity Emulation}
\label{subsec:mfemu_emu}

We briefly recap the multi-fidelity emulation framework we proposed in \cite{Ho:2022}.
We will first review the Kennedy-O'Hagan model (autoregressive GP; AR1) \citep{Kennedy:2000} and NARGP (non-linear autoregressive GP) \citep{Perdikaris:2017} in Section~\ref{subsubsec:mfemu_emu_ko} and Section~\ref{subsubsection:nargp}, respectively.
We do not change our AR1 and NARGP modeling presented in \cite{Ho:2022}, except we simplified the notations to only two fidelities.
Finally, we will introduce the GMGP model \citep{Ji:2021}, combining simulations from different box sizes.

\subsubsection{Kennedy O'Hagan Method}
\label{subsubsec:mfemu_emu_ko}

\cite{Kennedy:2000} proposed a linear autoregressive GP to model the response surfaces of a sequence of computer codes with increasing fidelity.
For simplicity, we assume there are only two fidelities:
dark-matter only simulations with fewer particles in low fidelity (LF)
and with more particles in high fidelity (HF).

Let $\{\outputVector_\mathrm{LF}, \outputVector_\mathrm{HF}\}$ be the matter power spectrum in the training set, where $\outputVector_\mathrm{LF} = \{\outputFunction_\mathrm{LF}(\thetavec^\mathrm{LF}_i) \}_{i=1}^{n_\mathrm{LF}}$  and $\outputVector_\mathrm{HF} = \{\outputFunction_\mathrm{HF}(\thetavec^\mathrm{HF}_i) \}_{i=1}^{n_\mathrm{HF}}$. Here $n_\mathrm{LF}$ and $n_\mathrm{HF}$ are the number of simulations in the low and high fidelity.
The KO method models the multi-fidelity emulator as:
\begin{equation}
    \outputFunction_\mathrm{HF}(\thetavec) = \rho \cdot \outputFunction_\mathrm{LF}(\thetavec)
    + \delta(\thetavec),
    \label{eq:ko_model}
\end{equation}
where $\rho$ (the scale parameter) is a trainable parameter describing the amount of common behavior in low- and high-fidelity response surfaces.
$\delta(\thetavec)$ is a GP that models the remaining bias, modeling the variability that cannot be captured by correlating LF to HF.
In the context of the matter power spectrum, the
$\rho \cdot \outputFunction_\mathrm{LF}(\thetavec)$ term dominates at the large scales describing the two-halo term while $\delta(\thetavec)$ dominates at the small scales describing the one-halo term.

We normalize the matter power spectra into a logarithmic scale.
The sample mean is subtracted from the LF log power spectra to keep the output close to zero, while the HF log power spectra are passed directly to the training:
\begin{equation}
    \begin{split}
        \yvec_\mathrm{LF} &\leftarrow \log{\yvec_\mathrm{LF}} - \mathbb{E}[\log{\yvec_\mathrm{LF}}] ;\\
        \yvec_\mathrm{HF} &\leftarrow \log{\yvec_\mathrm{HF}}.
    \end{split}
    \label{eq:normalization}
\end{equation}
Not subtracting the mean spectrum of HF simulations is a compromise decision.
Our benchmark multi-fidelity emulator uses only 3 HF samples, and the sample mean of 3 power spectra will often deviate substantially from the true mean spectrum.
Instead, we entirely rely on the bias term, $\delta(\thetavec)$, to compensate for the deviation caused by not subtracting the mean.


As mentioned in \cite{Ho:2022},
the $\rho$ parameter has to be scale-dependent (as a function of $k$) to model the scale-dependent correlation between high- and low-fidelity.
Here we use the same method as \cite{Ho:2022}, where we assume Equation~\ref{eq:ko_model} is a single-output GP model and build a KO model for each $k$ bin of the data.
In this way, we can model $\rho$ as a function of $k$.


We also assign different KO models to different redshifts.
We note that it is possible to assume a smooth function to model $\rho(k, z)$,
and we may examine this in future work.

\subsubsection{Non-linear Autoregressive Gaussian Process (NARGP)}
\label{subsubsection:nargp}

Another multi-fidelity method we used in \cite{Ho:2022} is the non-linear autoregressive GP, or NARGP, developed by \cite{Perdikaris:2017}.
NARGP is a modification of the KO method to allow non-linearity in the scale parameter, $\rho$, through a deep GP \citep{Damianou:2013}.
In cosmic emulators, it means that we allow $\rho$ to vary as a function of cosmology.

Let $\outputFunction_\mathrm{HF}(\thetavec)$ be the high-fidelity and $\outputFunction_\mathrm{LF}(\thetavec)$ be the low-fidelity power spectra as functions of cosmology, $\thetavec$.
NARGP models the multi-fidelity problem as:
\begin{equation}
    \outputFunction_\mathrm{HF}(\thetavec) = \rho(\thetavec,  \outputFunction_\mathrm{LF}(\thetavec))
    + \delta(\thetavec),
    \label{eq:nargp}
\end{equation}
Here, $\rho$ is modeled as a GP and is a function of the cosmologies, $\thetavec$, and the output from the previous fidelity, $\outputFunction_\mathrm{LF}(\thetavec)$.
We follow the approximation made in \cite{Perdikaris:2017} to simplify the computation of a deep GP to two separate GPs.
The approximation is done by replacing the $\outputFunction_\mathrm{LF}(\thetavec)$ with its posterior, $\outputFunction_\mathrm{*,LF}(\thetavec)$.
Eq~\ref{eq:nargp} can thus be further reduced to a regular GP with a kernel function $K$:
\begin{equation}
    \outputFunction_\mathrm{HF} \sim \GP(0, K)
    \label{eq:nargp_posterior}
\end{equation}
with
\begin{equation}
    K(\thetavec, \thetavec') =
    K_\rho(\thetavec, \thetavec') \cdot
    K_\outputFunction(\outputFunction_\mathrm{*,LF}(\thetavec), \outputFunction_\mathrm{*,LF}'(\thetavec'))
    +
    K_\delta(\thetavec, \thetavec').
    \label{eq:nargp_kernel}
\end{equation}
We integrate the bias GP and the scale parameter GP here into one single GP with a composite kernel.
Each kernel, $(K_\rho, K_\outputFunction, K_\delta)$, is a squared exponential kernel.
$K_\delta$ models the bias term, and the scale parameter GP is factorized into the $K_\outputFunction$, modeling the covariance between LF output posteriors. $K_\rho$ models the cosmological dependence of $\rho$.

\subsubsection{Graphical Multi-fidelity Gaussian Process (GMGP)}
\label{subsubsec:mfemu_emu_gmgp}

Here we briefly explain a new multi-fidelity model using a graphical model Gaussian process (GMGP), first introduced in \cite{Ji:2021}.
A similar approach is the multi-information source method \citep{MISO:2016}, which allows multiple low-fidelity nodes (information sources) to resolve a single high-fidelity truth.
However, we find the model in \cite{Ji:2021} is methodologically closer to what we applied before in \cite{Ho:2022},
and so use this technique for our emulation problem for low-fidelity nodes with different box sizes.

The graphical GP model \citep{Ji:2021} utilizes a directed acyclic graph to model multi-fidelity data.
Instead of assuming the fidelities of a simulation code form a monotonically increasing sequence in accuracy,
a GMGP allows the fidelities to have a directed-in tree structure. \cite{Ji:2021} has a thorough mathematical description for applying GMGP in an arbitrarily directed in-tree structure.
Thus each high fidelity node has more than one corresponding low fidelity node, a common situation as there are many ways to approximate high fidelity simulations.

We use the simplest case of the tree structure, illustrated in Fig~\ref{fig:nbody_plot}, with two low fidelity nodes and one high fidelity node.
In the case of $N$-body simulations, one may vary not only the number of particles, but also the box size of the simulation.
Thus we can use a low-fidelity simulation with a smaller box size to improve emulation at the high-fidelity node. We will call this tree ``{\mfbox}'' throughout the rest of the paper.
In the following text, we will assume L1 is the low-fidelity node that has $128^3$ particles. L2 has the same number of particles as L1 but a smaller box size ($100 \Mpch$), and HF is the high-fidelity node with $512^3$ particles and the same box size as L1 (Table~\ref{table:simulations}).


The deep GMGP model (dGMGP) we use from \cite{Ji:2021} is an extension of NARGP, where \cite{Ji:2021} implemented a specific kernel structure allowing low-fidelity information from multiple nodes to be passed to the HF node\footnote{Since we found NARGP outperformed AR1 in \cite{Ho:2022} for the matter power spectrum case, we will use dGMGP instead of the GMGP extended from the AR1 model.}.
For the directed graph in Fig~\ref{fig:nbody_plot},
the dGMGP model can be written as:
\begin{equation}
    \begin{split}
        \outputFunction_\mathrm{HF}(\thetavec) &= \rho(\{ \outputFunction_\mathrm{t}(\thetavec) : t \in L1, L2 \}, \thetavec) + \delta(\thetavec).
    \end{split}
    \label{eq:dgmgp}
\end{equation}
Here we pass the cosmologies $\thetavec$ and the outputs from L1 and L2 to the $\rho$ function.
We make the same approximation as in Section~\ref{subsubsection:nargp},
so we can train the deep GP recursively:
We first train the low-fidelity emulators on L1 and L2, respectively.
Then, we sample the output posteriors from the L1 and L2 emulators and use them as the training input for Eq~\ref{eq:dgmgp}.

Similar to NARGP, we use a composite kernel for the high-fidelity GP in the dGMGP:
\begin{equation}
    \begin{split}
        K_\mathrm{dGMGP}(\thetavec, \thetavec') &=\\
        K_\rho(\thetavec, \thetavec')
        &\cdot
        K_\outputFunction(\outputFunction_\mathrm{*,LF}(\thetavec), \outputFunction_\mathrm{*,LF}(\thetavec'))
        +
        K_\delta(\thetavec, \thetavec'),
    \end{split}
    \label{eq:dgmgp_kernel}
\end{equation}
where the above expression is the same as Eq~\ref{eq:nargp_kernel} except that $K_f$ takes the outputs from both L1 and L2 emulators as inputs,
\begin{equation}
    \begin{split}
        K_\outputFunction(&\outputFunction_\mathrm{*,LF}(\thetavec), \outputFunction_\mathrm{*,LF}(\thetavec'))
        =\\
        &K_\mathrm{linear}(\outputFunction_\mathrm{*,LF}(\thetavec), \outputFunction_\mathrm{*,LF}(\thetavec'))
        +\\
        &K_\mathrm{rbf}(\outputFunction_\mathrm{*,L1}(\thetavec), \outputFunction_\mathrm{*,L1}(\thetavec')) \cdot
        K_\mathrm{rbf}(\outputFunction_\mathrm{*,L2}(\thetavec), \outputFunction_\mathrm{*,L2}(\thetavec')).
    \end{split}
    \label{eq:dgmgp_kernel_decompose}
\end{equation}
Here, $K_\mathrm{rbf}$ is a radial basis kernel, and $K_\mathrm{linear}$ is a linear kernel, which can be expressed more explicitly as
\begin{equation*}
    K_\mathrm{linear}(\outputFunction_\mathrm{*,LF}, \outputFunction_\mathrm{*,LF}')
    = \sigma_1^2 \outputFunction_\mathrm{*,L1} \outputFunction_\mathrm{*,L1}'
    + \sigma_2^2 \outputFunction_\mathrm{*,L2} \outputFunction_\mathrm{*,L2}',
\end{equation*}
where $\sigma_1^2$ and $\sigma_2^2$ are the hyperparameters of the linear kernel.
A linear kernel in a Gaussian process is equivalent to a Bayesian linear regression.\footnote{See the kernel cookbook: \url{https://www.cs.toronto.edu/~duvenaud/cookbook/}.}
The multiplication in the kernel operation means an ``AND'' operation, showing high covariance only if both kernels have high values.
The addition operator means an ``OR'' operation, indicating the final covariance is high if either of the kernels gives a high value.
The intuition here is that the linear kernel encodes the linear regression part while the multiplication of RBF kernels encodes the non-linear transformation from L1 and L2 nodes to the HF node

\section{Sampling strategy for high-fidelity simulations}
\label{sec:sampling}

This section describes the method used for selecting the input parameters for our high-fidelity training simulations.
Following \cite{Ji:2021}, we employ a Sliced Latin Hypercube Design (SLHD) \citep{Qian:2012,Ba:2015} to assign input parameters for the high-fidelity (HF) nodes.
Each slice (or subset) in an SLHD is a Latin hypercube
and thus can be served as the design points for the HF node.
This approach offers a less computationally intensive and more straightforward implementation compared to  the grid search method utilized in our previous work  \citep{Ho:2022}.
The details of SLHD will be discussed in Section~\ref{subsec:slhd}, and our process for selecting the optimal HF design from the SLHD will be discussed in Section~\ref{subsec:optimal_slice}.

\subsection{Sliced Latin hypercube design (SLHD)}
\label{subsec:slhd}

Sliced Latin Hypercube Design (SLHD) is a type of Latin hypercube that can be partitioned by slices or blocks, each of which contains an equal number of design points. Each slice is itself a Latin hypercube.
SLHD ensures the space-filling property both in the whole design and in each slice.
Therefore, SLHD is an intuitive choice for a multi-fidelity problem.

Suppose we have an SLHD for the LF node.
We can use one of the slices to generate simulations for the HF node, which ensures that both the LF and HF nodes are in Latin hypercubes.
Another advantage of SLHD is that we can directly obtain a nested experimental design where the LF samples form a superset of the HF samples, i.e., $\thetavec_\mathrm{HF} \subset \thetavec_\mathrm{LF}$.
As mentioned in \cite{Kennedy:2000}, a nested design is an efficient training set for a multi-fidelity model because it allows us to obtain an accurate posterior $f_\mathrm{LF}(\theta)$ at location $\theta$ without interpolating at the low fidelity.

SLHD, initially proposed by \cite{Qian:2012}, is a technique developed for applying the Latin hypercube design to categorical variables.
\cite{Ba:2015} later developed an efficient method for constructing optimal SLHD designs.
The number of categories for categorical variables is usually fixed based on qualitative properties, making it challenging to apply a Latin hypercube design to such variables.
However, SLHD addresses this challenge and enables the use of Latin hypercube designs with categorical variables.
In SLHD, a Latin hypercube is divided into equal slices along the dimensions associated with categorical variables, while non-categorical dimensions are still sampled with ordinary Latin hypercube sampling.
The usage of SLHD in the context of modeling the multi-fidelity problem was demonstrated in \cite{Ji:2021}.
Furthermore, SLHD has also been employed in cosmology, specifically by the Dark Emulator \citep{Nishimichi:2019}.

For implementation,
we use the maximin SLHD package, \texttt{maxminSLHD},\footnote{\url{https://rdrr.io/cran/SLHD/man/maximinSLHD.html}} in \texttt{R} \citep{Ba:2015}.
We set the number of design points to $3$ for each slice and the number of slices to $20$.
In total, we have $60$ design points.
We assign the SLHD with $60$ points to LF and select one slice as our HF design.
We use $60$ LF points in this work because we learned in \cite{Ho:2022} that $\sim 50$ simulations are enough for a $5$ dimensional emulation problem.

\subsection{Selecting the optimal slice}
\label{subsec:optimal_slice}

Slices in SLHD are Latin hypercubes in smaller sizes.
In principle any slice should produce reasonably good emulation, as the points in each slices span parameter space.

\begin{figure}
    \includegraphics[width=\columnwidth]{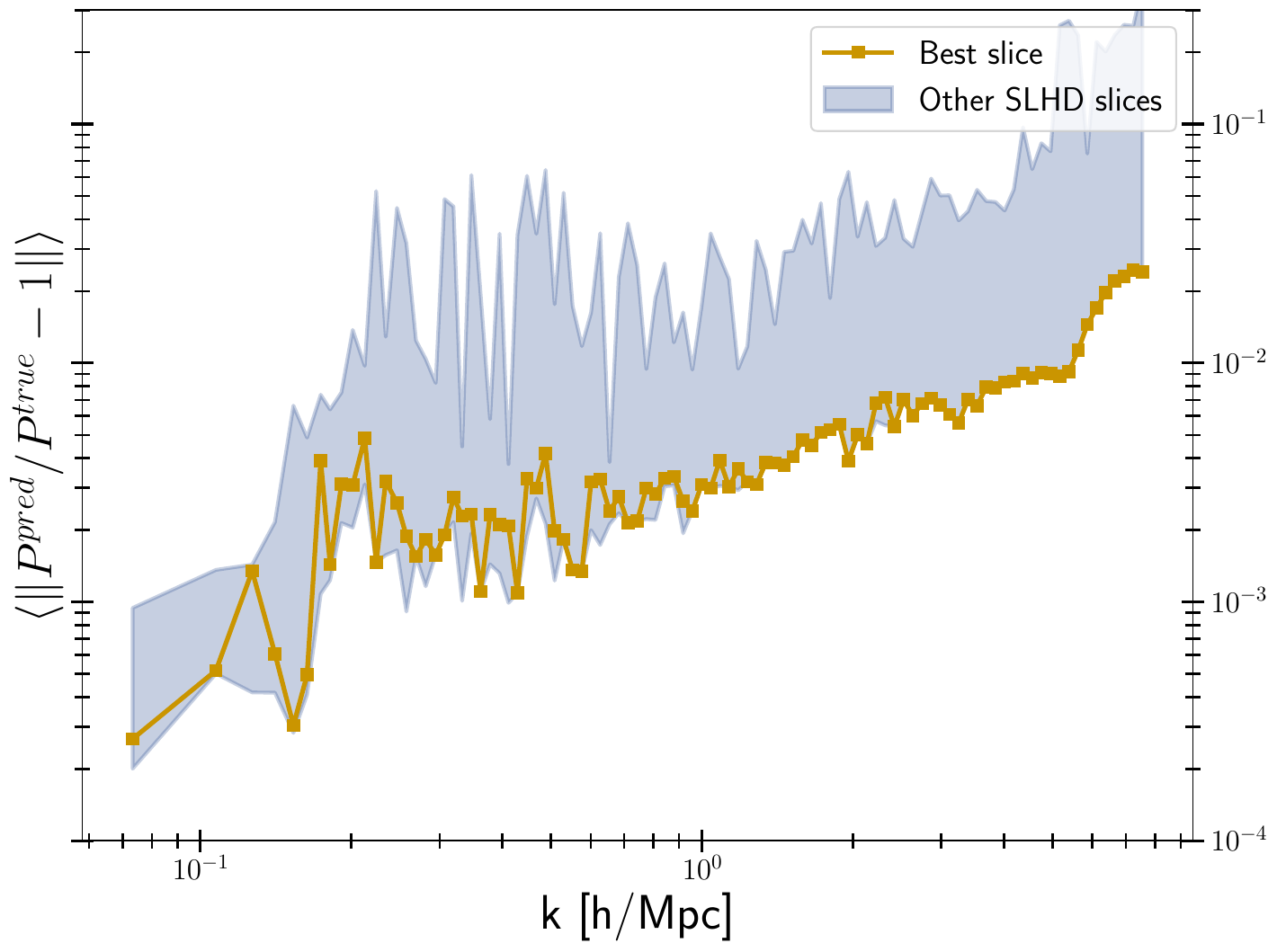}
    \caption{{\mfbox}'s emulation errors, averaged over redshift bins and test simulations, using 60 L1, 60 L2, and 3 HF (see Table~\ref{table:simulations}).
    Here, we show the emulation minimum and maximum errors using different slices from SLHD (\textcolor{skyline}{blue shaded area}), and the best slice found by the grid search method is labeled as \textcolor{sunshine}{yellow}.
    }
    \label{fig:slhd_slices}
\end{figure}

However, in practice some slices still perform somewhat better than others, as shown in Figure~\ref{fig:slhd_slices}. We use a procedure similar to our grid search approach in \cite{Ho:2022} to avoid choosing the worst slice.
The procedure is described below:
\begin{enumerate}
    \item Prepare SLHD for LF simulation suite.
    \item Build low-fidelity only emulators (LFEmu) for each slice, compute the interpolation error for each LFEmu, testing solely on the LF simulation suite.
    \item Select the slice which can best minimize the interpolation error.
\end{enumerate}

Note that we do not use any HF simulations in the above procedure.
The selection entirely relies on the LF simulation suite.
The underlying assumption is that the interpolation error of the low-fidelity node is correlated with the interpolation error of the high-fidelity node.
We labeled the selected slice in Figure~\ref{fig:LHS_samples}.
We will use the best slice as our HF training set for the results in Section~\ref{sec:results}.

To summarize, SLHD is a special kind of LHD, with each slice in the SLHD being a Latin hypercube as well as the whole design.
We thus can assign HF nodes with a slice (or slices) of SLHD, making both LF and HF nodes Latin hypercubes.
In the end, we describe a procedure to avoid choosing the worst slice for training a {\mfemu}.

\section{Computational budget estimation}
\label{sec:budget}


In this section, we present our approach to quantifying the optimal allocation of simulation budgets across different fidelities.
Building upon the error bounds established in \cite{Ji:2021}, we have made modifications to adapt them to our specific context, as described in Section~\ref{subsec:error_bounds}.
We approximate the emulation errors of our {\mfbox} using the form of \cite{Ji:2021} and empirically infer the error function of the emulator for various training designs, denoted as $(n_\mathrm{L1}, n_\mathrm{L2}, n_\mathrm{HF})$. Our objective is to utilize this empirical error function to determine the most cost-effective strategy for assigning low- and high-fidelity simulations in order to achieve optimal accuracy.

In Section~\ref{subsec:error_bounds}, we present an approximate error function for our {\mfbox} emulator in predicting high-fidelity simulation outputs.
Next, in Section~\ref{subsec:optimal_design}, we show the analysis for assigning optimal computational budgets to low- and high-fidelity simulations, under the assumption that the emulator error follows the approximate error function.
In Section~\ref{subsec:error_empirical}, we empirically estimate the approximate error function of the {\mfbox} by analyzing the average emulator errors obtained from 144 distinct {\mfbox} training results.
Finally, we determine the optimal number of low- and high-fidelity simulations required for achieving accurate power spectra emulation using the {\mfbox} approach.

\subsection{Error bounds for Gaussian process emulators}
\label{subsec:error_bounds}

\cite{Ji:2021} presents an error bound for a multi-fidelity emulator, and for the case of two low-fidelity nodes, the form is given by $\sim \mathcal{O}(\rho_\mathrm{L1} \cdot n_\mathrm{L1}^{-\frac{\nu_\mathrm{L1}}{d}} +
\rho_\mathrm{L2} \cdot n_\mathrm{L2}^{-\frac{\nu_\mathrm{L2}}{d}} +
n_\mathrm{HF}^{-\frac{\nu_\mathrm{HF}}{d}})$,
where $(\rho_\mathrm{L1}, \rho_\mathrm{L2})$ are the scale parameters for the L1 and L2 nodes, respectively.
$(\nu_\mathrm{L1}, \nu_\mathrm{L2}, \nu_\mathrm{HF})$ are positive spectral indices, and $(n_\mathrm{L1}, n_\mathrm{L2}, n_\mathrm{HF})$ represent the number of training simulations at the L1, L2, and HF nodes, respectively.
While this bound does not directly apply to our case, we utilize the form of the bound as an approximate model for the {\mfbox} error and empirically determine the parameters by fitting them to the {\mfbox} emulation results using different multi-fidelity designs, i.e., varying combinations of $(n_\mathrm{L1}, n_\mathrm{L2}, n_\mathrm{HF})$.

The equation below represents the error function of the {\mfbox} emulator we want to infer. Note that our discussion primarily focuses on the emulation error when predicting ``high-fidelity'' power spectra. This emphasis aligns with the core objective of {\mfbox}, which is to correct the resolution of low-fidelity simulations for accurate predictions of their high-fidelity counterparts.
\begin{equation}
    \begin{split}
        \Phi(n_\mathrm{L1}, &n_\mathrm{L2}, n_\mathrm{HF})=
        \frac{1}{N} \sum_{i = 1}^N
        \left\lvert
        \frac{
            \outputFunction_\mathrm{HF}(\thetavec_i) - m_{\outputFunction_\mathrm{HF}}(\thetavec_i)
        }{\outputFunction_\mathrm{HF}(\thetavec_i)}
        \right\rvert\\
        &\approx
        \tilde\Phi(n_\mathrm{L1}, n_\mathrm{L2}, n_\mathrm{HF})\\
        &=
        \eta \cdot (
            \rho_\mathrm{L1} \cdot n_\mathrm{L1}^{- \frac{\nu_\mathrm{L1}}{d}} +
            \rho_\mathrm{L2} \cdot n_\mathrm{L2}^{- \frac{\nu_\mathrm{L2}}{d}} +
            n_\mathrm{HF}^{- \frac{\nu_\mathrm{HF}}{d}}
        ),
    \end{split}
    \label{eq:error_function_approx}
\end{equation}
where $N = 10$ test simulations in a Latin hypercube are used to average the emulation relative error.
The emulator error function $\Phi(n_\mathrm{L1}, n_\mathrm{L2}, n_\mathrm{HF})$ represents the average relative error of the {\mfbox} as a function of the number of simulations in L1, L2, and HF nodes.
To estimate this error function, we have already averaged the emulation error across $k$ bins, enabling us to obtain an approximation of the error as a function of the design points ($n_\mathrm{L1}, n_\mathrm{L2}, n_\mathrm{HF}$).
Then, we infer the parameters of this error function from the {\mfbox} emulation results, as denoted by the $\approx$ sign in Eq~\ref{eq:error_function_approx}.
The normalization factor of the functional form in Eq~\ref{eq:error_function_approx} is determined by the free parameter $\eta$.

An important term in Eq~\ref{eq:error_function_approx} is the one describing how the error scales with an increasing number of simulations, $n_t^{-\frac{1}{d}}$, where $t \in {\mathrm{L1}, \mathrm{L2}, \mathrm{HF}}$.
This scaling term comes from the fact that the fill distance is proportional to $\mathcal{O}(n_t^{-\frac{1}{d}})$, where $d$ is the number of dimensions in a space-filling design \citep{Wendland:2004}.

\begin{figure}
    \includegraphics[width=\columnwidth]{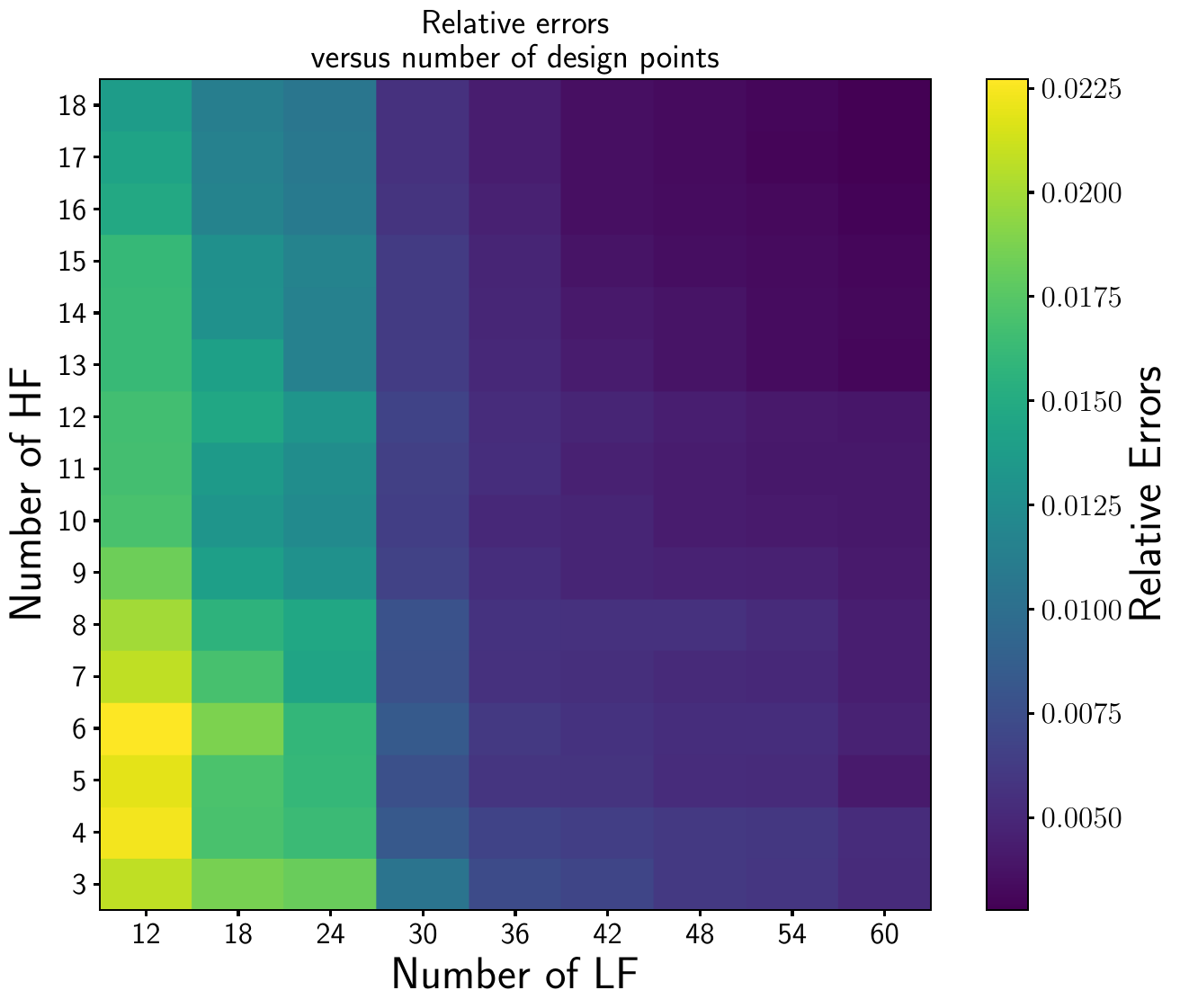}
    \caption{Relative errors plotted against the number of LF and HF design points in a {\mfbox} emulator.
    Here, LF refers to the combined number of L1 and L2 points, where LF $= n_\mathrm{L1} = n_\mathrm{L2}$.
    The plot reveals a trend of decreasing errors as the number of low-fidelity training simulations increases.
    However, due to the limited number of high-fidelity points compared to LF points, the decreasing trend is relatively modest.
    }
    \label{fig:budget_imshow_errors}
\end{figure}

To determine the parameters of $\tilde\Phi(n_\mathrm{L1}, n_\mathrm{L2}, n_\mathrm{HF})$, we employ Markov Chain Monte Carlo (MCMC) inference based on 144 distinct {\mfbox} emulators that were trained with varying numbers of $(n_\mathrm{L1}, n_\mathrm{L2}, n_\mathrm{HF})$.
Specifically, we generated {\mfbox} emulators using $[12, 18, 24, \cdots, 60]$ L1/L2 points and $[2, 3, \cdots, 18]$ HF points, resulting in a total of 144 emulators.
For simplicity, we only considered cases where the number of simulations in L1 and L2 nodes was equal, i.e., $n_\mathrm{L1} = n_\mathrm{L2}$, as the costs of L1 and L2 nodes are similar, therefore, choosing between them is not important.
To simplify the notation, we employ $n_\mathrm{LF}$ to represent the number of training points in both the L1 and L2 nodes.
Figure~\ref{fig:budget_imshow_errors} presents the average relative errors, $\Phi(n_\mathrm{L1}, n_\mathrm{L2}, n_\mathrm{HF})$, for all 144 designs under consideration.

For each pixel in Figure~\ref{fig:budget_imshow_errors}, we compute the average emulator relative error across $10$ test simulations and multiple $k$ bins across a redshift range, $z \in [0, 0.2, 0.5, 1, 2, 3]$.
To solve the parameter estimation problem, we employ Markov Chain Monte Carlo (MCMC) inference with a Gaussian likelihood,
\footnote{We use the \texttt{PyMC} package version 4 \citep{Salvatier:2016} for the MCMC inference.}
\begin{equation}
    \begin{split}
        \tilde\Phi(&n_\mathrm{L1}, n_\mathrm{L2}, n_\mathrm{HF})\\
        &=
        \eta \cdot (
            \rho_\mathrm{L1} \cdot n_\mathrm{L1}^{- \frac{\nu_\mathrm{L1}}{d}} +
            \rho_\mathrm{L2} \cdot n_\mathrm{L2}^{- \frac{\nu_\mathrm{L2}}{d}} +
            n_\mathrm{HF}^{- \frac{\nu_\mathrm{HF}}{d}}
        )\\
        &\sim
        \normal(
            \mu=\Phi(n_\mathrm{L1}, n_\mathrm{L2}, n_\mathrm{HF}),
            \sigma^2=\Phi_\mathrm{var}(n_\mathrm{L1}, n_\mathrm{L2}, n_\mathrm{HF})
        ).
    \end{split}
    \label{eq:error_function_likelihood}
\end{equation}
Here, $\Phi(n_\mathrm{L1}, n_\mathrm{L2}, n_\mathrm{HF})$ represents the average relative errors, while $\Phi_\mathrm{var}(n_\mathrm{L1}, n_\mathrm{L2}, n_\mathrm{HF})$ denotes the variance of the relative errors across 10 test simulations.

\begin{table*}
	\centering
	\caption{MCMC analysis of Eq~\protect\ref{eq:error_function_approx}:
    $\frac{1}{N} \sum_{i = 1}^N
    \left\lvert
    \frac{
        \outputFunction_\mathrm{HF}(\thetavec_i) - m_{\outputFunction_\mathrm{HF}}(\thetavec_i)
    }{\outputFunction_\mathrm{HF}(\thetavec_i)}
    \right\rvert
    =
    \Phi(n_\mathrm{L1}, n_\mathrm{L2}, n_\mathrm{HF})
    \approx
    \eta \cdot (
    \rho_\mathrm{L1} \cdot n_\mathrm{L1}^{- \frac{\nu_\mathrm{L1}}{d}} +
    \rho_\mathrm{L2} \cdot n_\mathrm{L2}^{- \frac{\nu_\mathrm{L2}}{d}} +
    n_\mathrm{HF}^{- \frac{\nu_\mathrm{HF}}{d}}).$
    The notation $\{\Phi(n_{\mathrm{L1},j}, n_{\mathrm{L2},j}, n_{\mathrm{HF},j})\}_{j=1}^{144}$ means all 144 {\mfbox} emulator errors used for parameter estimation.
    The column ``Posterior ($50\%$)'' reports the medians of the posteriors of the parameters, and ``Posterior ($25\%$, $75\%$)'' reports the 25\% and 75\% quantities of the posterior distributions.
    }
	\label{tab:budget_size_mcmc}
	\begin{tabular}{lcccc}
		\hline
		Parameters & Prior & Posterior ($50\%$) & Posterior ($25\%$, $75\%$)\\
		\hline
		$\eta$ & $\mathrm{Normal}(\mu = \mathrm{Mean}(\{\Phi_j\}_{j=1}^{144}), \sigma^2 = \mathrm{Var}(\{\Phi_j\}_{j=1}^{144}))$ & $0.0308$ &  $(0.0290, 0.0327)$   \\
		$\nu_\mathrm{HF}$ & $\mathrm{LogNormal}(\mu = 0, \sigma = 1)$ & $9.80$   &  $(9.44, 10.2)$ \\
        $\nu_\mathrm{L1}$ & $\mathrm{LogNormal}(\mu = 0, \sigma = 1)$ & $5.49$   &  $(5.33, 5.67)$ \\
		$\nu_\mathrm{L2}$ & $\mathrm{LogNormal}(\mu = 0, \sigma = 1)$ & $5.49$ &  $(5.33, 5.67)$ \\
        $\rho_\mathrm{L1}$ & $\mathrm{Normal}(\mu = 1, \sigma = 1)$ & $4.53$ &  $(3.97, 5.08)$ \\
        $\rho_\mathrm{L2}$ & $\mathrm{Normal}(\mu = 1, \sigma = 1)$ & $4.54$ &  $(3.97, 5.10)$ \\
		\hline
	\end{tabular}
\end{table*}

The results of our MCMC analysis, including the priors and posteriors, are summarized in Table~\ref{tab:budget_size_mcmc}.
The posteriors show that $\nu_\mathrm{L1} \simeq \nu_\mathrm{L2}$ and $\rho_\mathrm{L1} \simeq \rho_\mathrm{L2}$, indicating that both L1 and L2 nodes contribute to improving the accuracy of the emulator in a similar manner.
In contrast, the power-law index $\nu_\mathrm{HF}$ for the HF node is approximately twice as large as $\nu_\mathrm{L1}$ and $\nu_\mathrm{L2}$, suggesting that the HF node has a more pronounced impact on enhancing the emulator's accuracy compared to the LF nodes.
Table~\ref{tab:budget_size_mcmc} shows that the parameters in Eq~\ref{eq:error_function_approx} are reasonably well-defined.
Thus, we will use the median of the posterior as point estimates for the error function for the remainder of this paper.


\subsection{Optimal number of simulations per node}
\label{subsec:optimal_design}

Eq~\ref{eq:error_function_approx} models the emulation error, $\Phi(n_\mathrm{L1}, n_\mathrm{L2}, n_\mathrm{HF})$, which behaves as a combination of power-law functions of the number of simulations in each node, namely LF or HF.
The primary goal of an emulator is to better represent the original simulator by minimizing the prediction error, subject to a limited computational budget, denoted by $C$.
By using $\Phi(n_\mathrm{L1}, n_\mathrm{L2}, n_\mathrm{HF})$, we can determine the optimal number of simulations per node, given the computational budget available for running each node.

Consider a two-fidelity emulator consisting of two low-fidelity nodes, L1 and L2, where $\rho_\mathrm{L1,L2}$ are the scale parameters and $(n_\mathrm{L1}, n_\mathrm{L2}, n_\mathrm{HF})$ represent the number of simulations in L1, L2, and HF nodes, respectively.
Our goal is to minimize the emulation error while subject to a limited budget.
\begin{equation}
    n_\mathrm{L1} \cdot C_\mathrm{L1} +
    n_\mathrm{L2} \cdot C_\mathrm{L2} +
    n_\mathrm{HF} \cdot C_\mathrm{HF} \leq C,
\end{equation}
where we know the ratios between the costs of HF and LF nodes (L1 and L2) are $\frac{C_\mathrm{HF}}{C_\mathrm{L1}} \simeq 140$ and $\frac{C_\mathrm{HF}}{C_\mathrm{L2}} \simeq 140/1.7$, from Table~\ref{table:simulations}.

The Lagrangian for optimizing the error subjecting to the cost is:
\begin{equation}
    \begin{split}
        \mathcal{L}(n_\mathrm{L1}, &n_\mathrm{L2}, n_\mathrm{HF}, \lambda) =
        \eta (
            \rho_\mathrm{L1} \cdot n_\mathrm{L1}^{- \frac{\nu_\mathrm{L1}}{d}}
            + \rho_\mathrm{L2} \cdot n_\mathrm{L2}^{- \frac{\nu_\mathrm{L2}}{d}}
            + n_\mathrm{HF}^{- \frac{\nu_\mathrm{HF}}{d}})\\
        &+
        \lambda (
            n_\mathrm{L1} \cdot C_\mathrm{L1}
            + n_\mathrm{L2} \cdot C_\mathrm{L2}
            + n_\mathrm{HF} \cdot C_\mathrm{HF} - C
            ),
    \end{split}
\end{equation}
Here, $\lambda$ is the Lagrange multiplier.
To find the optimal number of $(n_\mathrm{L1}, n_\mathrm{L2}, n_\mathrm{HF})$ minimizing the emulation error, we use the 1st order derivative conditions of the Lagrangian,
\begin{equation}
    \begin{split}
        \frac{\partial \mathcal{L}(n_\mathrm{L1}, n_\mathrm{L2}, n_\mathrm{HF}, \lambda)}{\partial n_\mathrm{L1}} = 0;\\
        \frac{\partial \mathcal{L}(n_\mathrm{L1}, n_\mathrm{L2}, n_\mathrm{HF}, \lambda)}{\partial n_\mathrm{L2}} = 0;\\
        \frac{\partial \mathcal{L}(n_\mathrm{L1}, n_\mathrm{L2}, n_\mathrm{HF}, \lambda)}{\partial n_\mathrm{HF}} = 0,
    \end{split}
\end{equation}
resulting in
\begin{equation}
    \begin{split}
        \eta \frac{\nu_\mathrm{L1}}{d} \rho_\mathrm{L1} \cdot n_\mathrm{L1}^{- \frac{\nu_\mathrm{L1} + d}{d}}
        =
        \lambda C_\mathrm{L1}
        &
        \Rightarrow
        n_\mathrm{L1} \propto (
            \frac{
                \nu_\mathrm{L1} \rho_\mathrm{L1}
            }{
                C_\mathrm{L1}
            })^{\frac{d}{\nu_\mathrm{L1} + d}} \\
        \eta \frac{\nu_\mathrm{L2}}{d} \rho_\mathrm{L2} \cdot n_\mathrm{L2}^{- \frac{\nu_\mathrm{L2} + d}{d}}
        =
        \lambda C_\mathrm{L2}
        &
        \Rightarrow
        n_\mathrm{L2} \propto (
            \frac{
                \nu_\mathrm{L2} \rho_\mathrm{L2}
            }{
                C_\mathrm{L2}
            })^{\frac{d}{\nu_\mathrm{L2} + d}} \\
        \eta \frac{\nu_\mathrm{HF}}{d} n_\mathrm{HF}^{- \frac{\nu_\mathrm{HF} + d}{d}}
        =
        \lambda C_\mathrm{HF}
        &
        \Rightarrow
        n_\mathrm{HF} \propto (
            \frac{
                \nu_\mathrm{HF}
            }{
                C_\mathrm{HF}
            })^{\frac{d}{\nu_\mathrm{HF} + d}}.
    \end{split}
    \label{eq:budget_n_nodes}
\end{equation}
Here, the intuition is relatively straightforward: the number of simulations required is inversely proportional to the cost of each simulation at a given fidelity.
However, if we observe a strong correlation between fidelities (i.e., if $\rho_\mathrm{L1,L2}$ is large), then we should use more low-fidelity simulations because they are less expensive.

To ensure that Eq~\ref{eq:budget_n_nodes} identifies local minima instead of maxima, we can verify the positivity of the second-order derivatives of the Lagrangian.
\begin{equation}
    \begin{split}
        \frac{\partial^2 \mathcal{L}(n_\mathrm{L1}, n_\mathrm{L2}, n_\mathrm{HF}, \lambda)}{\partial n_\mathrm{L1}^2}
        = \eta \rho_\mathrm{L1} \frac{\nu_\mathrm{L1} (\nu_\mathrm{L1} + d) }{d^2} n_\mathrm{L1}^{-\frac{\nu_\mathrm{L1} + 2d}{d}} > 0;\\
        \frac{\partial^2 \mathcal{L}(n_\mathrm{L1}, n_\mathrm{L2}, n_\mathrm{HF}, \lambda)}{\partial n_\mathrm{L2}^2}
        = \eta \rho_\mathrm{L2} \frac{\nu_\mathrm{L2} (\nu_\mathrm{L2} + d) }{d^2} n_\mathrm{L2}^{-\frac{\nu_\mathrm{L2} + 2d}{d}} > 0;\\
        \frac{\partial^2 \mathcal{L}(n_\mathrm{L1}, n_\mathrm{L2}, n_\mathrm{HF}, \lambda)}{\partial n_\mathrm{HF}^2}
        = \eta \frac{\nu_\mathrm{HF} (\nu_\mathrm{HF} + d) }{d^2} n_\mathrm{HF}^{-\frac{\nu_\mathrm{HF} + 2d}{d}} > 0.
    \end{split}
\end{equation}
The parameters $(\nu_\mathrm{L1}, \nu_\mathrm{L2}, \nu_\mathrm{HF})$, $(\rho_\mathrm{L1}, \rho_\mathrm{L2}, \rho_\mathrm{HF})$, and $\eta$ are all positive, while the dimension of the input space, $d$, must be a positive integer.
Similarly, the number of simulations $(n_\mathrm{L1}, n_\mathrm{L2}, n_\mathrm{HF})$ must be positive integers as well. Therefore, all second-order derivatives are positive, indicating that Eq~\ref{eq:budget_n_nodes} minimizes the emulation error.

In the special case where $\nu \equiv \nu_\mathrm{LF} = \nu_\mathrm{HF}$, Eq~\ref{eq:budget_n_nodes} simplifies to the optimal budget identified in \cite{Ji:2021}:
\begin{equation}
    \frac{n_\mathrm{LF}}{n_\mathrm{HF}}
    =
    \left(
        \frac{\rho_\mathrm{LF} C_\mathrm{HF}}{C_\mathrm{LF}}
    \right)^{\frac{d}{\nu + d}},
\end{equation}
where the ratio of LF/HF training sample sizes is inversely proportional to the cost of each simulation per run and directly proportional to the correlation with the high-fidelity node.

\begin{figure*}
    \includegraphics[width=2\columnwidth]{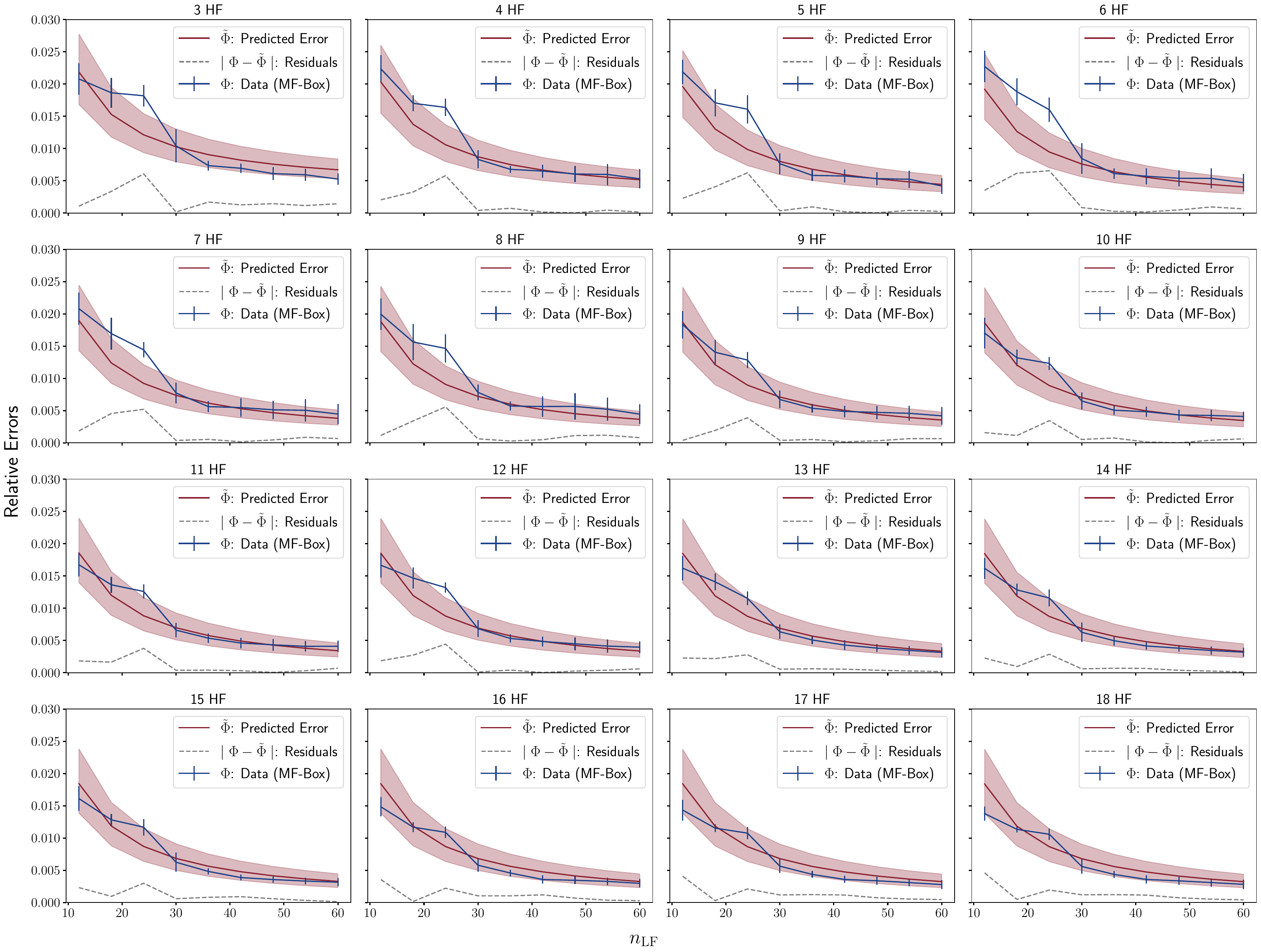}
    \caption{Inferred relative errors for all available {\mfbox} emulators are displayed.
    Each subplot corresponds to a fixed number of HF points (as indicated in the title) with varying LF points (on the x-axis).
    The red curves represent the median predictions (50\% posterior).
    Blue lines indicate the average relative errors obtained from the {\mfbox} emulators, while the error bars represent the standard deviation of relative errors across 10 simulations in the test set.
    The shaded area depicts the $25\%$ and $75\%$ confidence interval of the predictions based on the inference results.
    Overall, the relative errors demonstrate a decreasing trend as the number of LF and HF points increases.}
    \label{fig:budget_all_test_errors}
\end{figure*}

\subsection{Empirical estimate of the error function}
\label{subsec:error_empirical}

In this section, we present the predicted errors of {\mfbox} obtained from our MCMC analysis.
We explore the impact of different {\mfbox} designs on error predictions.
Finally, we discuss the choices of the optimal number of simulations for {\mfbox} based on the analysis presented in Section~\ref{subsec:optimal_design}.


We illustrate the predicted emulation errors in Fig~\ref{fig:budget_all_test_errors}, categorized by {\mfbox} models with varying LF and HF points.
The predictions align with the overall trend of the data, except when $n_\mathrm{LF}$ is low, where the limited availability of LF training points leads to suboptimal training performance.

\begin{figure}
    \includegraphics[width=\columnwidth]{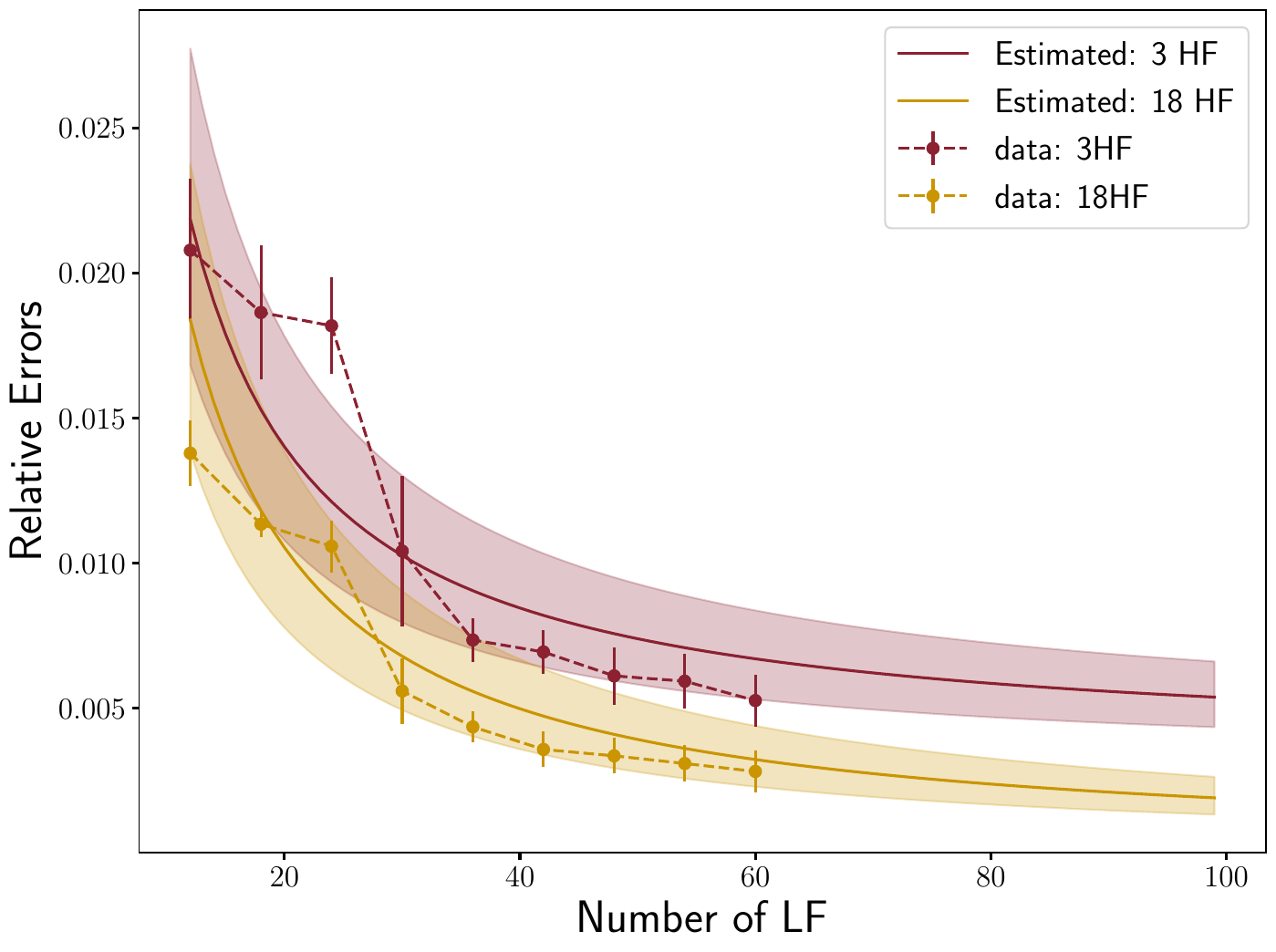}
    \caption{Inferred relative errors as a function of LF points.
    Shaded area shows the $25\%$ and $75\%$ confidence interval of the prediction from the inference result.}
    \label{fig:budget_inferred_LR_error}
\end{figure}

\begin{figure}
    \includegraphics[width=\columnwidth]{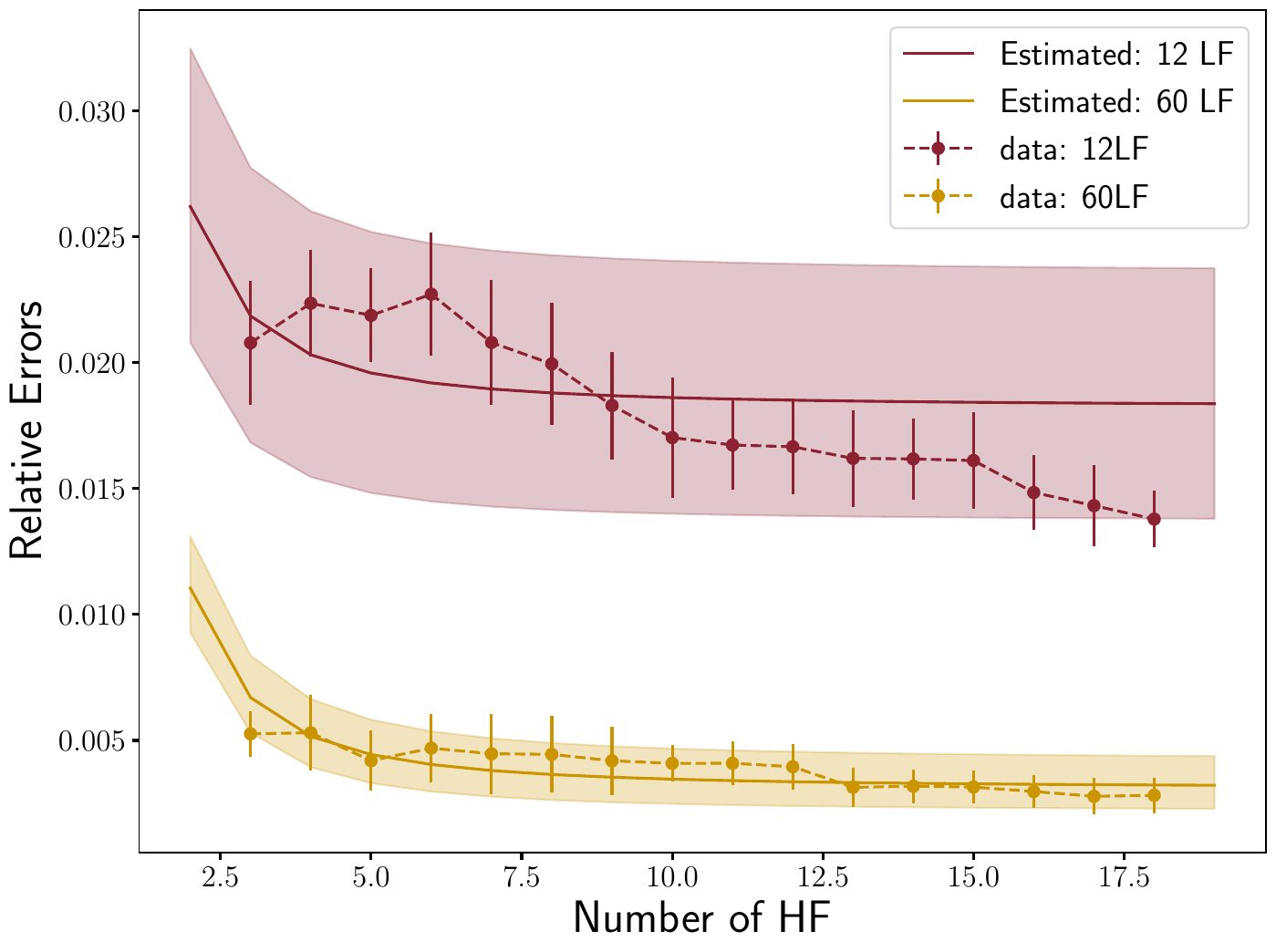}
    \caption{Inferred relative errors as a function of HF points.
    Shaded area shows the $25\%$ and $75\%$ confidence interval of the prediction from the inference result.}
    \label{fig:budget_inferred_HR_error}
\end{figure}

Fig~\ref{fig:budget_inferred_LR_error} and Fig~\ref{fig:budget_inferred_HR_error} depict the predicted relative errors as a function of LF and HF points, respectively.
Both figures exhibit a power-law trend characterized by a negative spectral index, indicating that the error decreases as the number of training points increases.
For example, in Fig~\ref{fig:budget_inferred_LR_error}, the \xyemulator{$X$}{3} emulators ($X \in \{12, 18, 24, 30, 36, 42, 48, 54, 60\}$) follow this trend concerning the number of LF points, suggesting that achieving further accuracy improvements becomes challenging once a sufficient number of LF points are used.
How much the error can be reduced by increasing the number of LF points is also influenced by the correlation between LF and HF simulations, which is controlled by the $\rho$ parameter.
A higher value of $\rho$ indicates that LF points can more effectively reduce the error.

On the other hand, incorporating additional HF points can also enhance accuracy.
In Fig~\ref{fig:budget_inferred_LR_error}, increasing the number of points in the HF node from 3 to 18 shifts the power-law function towards lower values, which itself follows the trend in Fig~\ref{fig:budget_inferred_HR_error}.
Similarly, as more HF points are included in the training, achieving further emulation accuracy becomes more challenging.

\begin{figure}
    \includegraphics[width=\columnwidth]{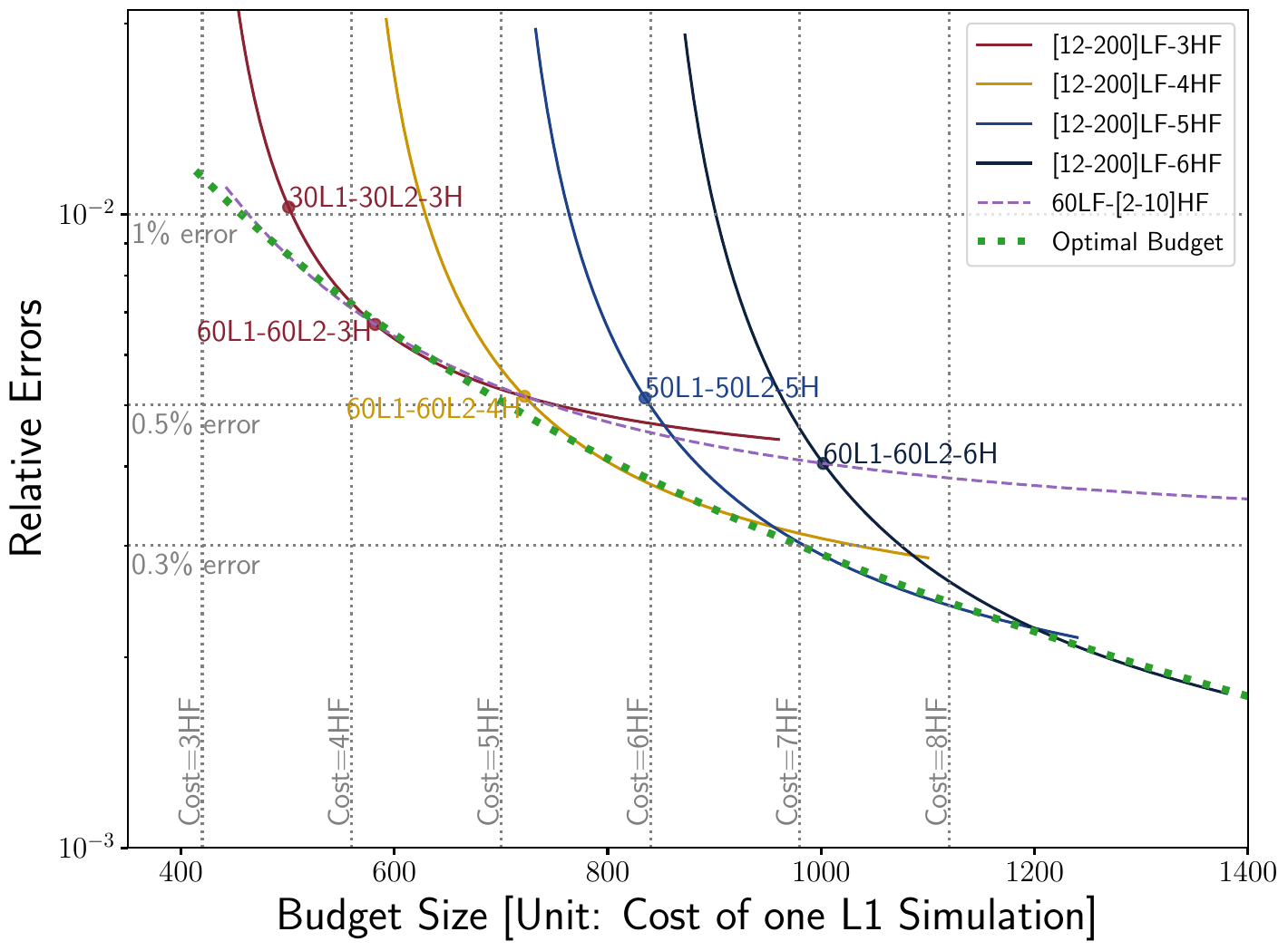}
    \caption{The predicted emulator errors as a function of the budget size, in the unit of the number of LF simulations.
    The predictions are based on the medians of the parameter posteriors presented in Table~\ref{tab:budget_size_mcmc}.
    The plot shows the predicted error functions using different combinations of LF and HF nodes.
    The red, yellow, blue, and black curves represent the predicted error functions with varying LF nodes and a fixed HF node ($n_\mathrm{HF} = 3, 4, 5, 6$).
    In contrast, the purple dashed curve represents the predicted error function with varying HF nodes and a fixed LF node ($n_\mathrm{LF} = 60$).
    The green dotted line illustrates the error function corresponding to the optimal budget (Eq~\ref{eq:budget_ratio}).
    The vertical gray dotted lines indicate the budget size in terms of the number of HF simulations.
    The horizontal gray dotted lines denote the predicted errors at the levels of $(1\%, 0.5\%, 0.3\%)$.
    }
    \label{fig:budget_all_costs_errors}
\end{figure}

Fig~\ref{fig:budget_all_costs_errors} displays the predicted error functions $\Phi(n_\mathrm{L1}, n_\mathrm{L2}, n_\mathrm{HF})$ for different {\mfbox} emulator designs.
We compile these predictions to create a plot of emulator error versus budget size.
The bottom left region of the plot represents the most economical budget setup, where the error is minimized relative to the allocated budget.

Based on the predictions in Figure~\ref{fig:budget_all_costs_errors}, we can determine the optimal number of simulations $(n_\mathrm{L1}, n_\mathrm{L2}, n_\mathrm{HF})$ for achieving a desired level of average accuracy.
For instance, if we aim for at least $1\%$ average error, the optimal choice is $(n_\mathrm{L1} = 30, n_\mathrm{L2} = 30, n_\mathrm{HF} = 3)$, which corresponds to a cost of approximately 500 L1 simulations.
Note that a minimum of 3 HF simulations ($\sim 420$ L1 simulations) is required to train a {\mfbox} in our power spectrum emulation problem.
Similarly, if we aim for at least $0.5\%$ average error, the optimal setup becomes $(n_\mathrm{L1} = 60, n_\mathrm{L2} = 60, n_\mathrm{HF} = 4)$. However, a slightly higher cost is required for the setup with $(n_\mathrm{L1} = 50, n_\mathrm{L2} = 50, n_\mathrm{HF} = 5)$, which yields a similar error.

In Figure~\ref{fig:budget_all_costs_errors}, the purple dashed curve represents the predicted error of \xyemulator{60}{[2-10]}s, illustrating the trend of increasing the number of HF points while keeping a fixed number of $60$ LF nodes.
At the point of (60 LF, 3 HF), the error decrease exhibits a similar gradient to \xyemulator{[12-200]}{3}, but it shows a steeper gradient after 4 HF points.
This result suggests that adding more LF or HF nodes does not necessarily lead to superior performance compared to each other.

Under the assumptions outlined in Section~\ref{subsec:optimal_design}, we can determine an optimal number of simulations $(n_\mathrm{L1}, n_\mathrm{L2}, n_\mathrm{HF})$ for a {\mfbox} to achieve the best emulation accuracy within a given computational budget.
The optimal ratio between the number of HF and LF simulations can be expressed as:
\begin{equation}
    \begin{split}
        n_\mathrm{LF}^{-\frac{\nu_\mathrm{LF} + d}{d}}
        =
        n_\mathrm{HF}^{-\frac{\nu_\mathrm{HF} + d}{d}}
        \frac{C_\mathrm{LF}}{C_\mathrm{HF}}
        \frac{
            \nu_\mathrm{HF}
        }{
            \rho_\mathrm{LF}
            \nu_\mathrm{LF}
        }\,.
    \end{split}
    \label{eq:budget_ratio}
\end{equation}
Here, LF is either L1 or L2.
$C_\mathrm{LF}$ and $C_\mathrm{HF}$ represent the computational cost of one simulation in the LF and HF, respectively.

In Figure~\ref{fig:budget_all_costs_errors}, the green dotted line represents the optimal budget according to Eq~\ref{eq:budget_ratio}.
When $n_\mathrm{HF} = 2.5$, the optimal number of low-fidelity simulations is $(n_\mathrm{L1}, n_\mathrm{L2}) = (80, 60)$, which is close to our initial setup of {\mfbox} with $(n_\mathrm{L1} = 60, n_\mathrm{L2} = 60, n_\mathrm{HF} = 3)$. 
Moreover, the design of $(n_\mathrm{L1} = 60, n_\mathrm{L2} = 60, n_\mathrm{HF} = 4)$ is also nearly optimal (close to the green dotted line), as demonstrated in Figure~\ref{fig:budget_all_costs_errors}.

In summary, this section introduces an approach to model the average emulation error of {\mfbox} as a function of LF and HF points using an approximate error model based on power-law functions.
Through empirical analysis of 144 {\mfbox} designs with various configurations, we have inferred this error model.
We demonstrate that this empirical model can guide the selection of an optimal design within a given computational budget, facilitating the construction of accurate emulators in a resource-efficient manner.

\section{Results}
\label{sec:results}

This section will demonstrate the emulation accuracy achieved by incorporating simulations with different box sizes through {\mfbox} for correcting the resolution of low-fidelity emulators to predict high-fidelity counterparts. The emulation error in this section is computed using a hold-out test set comprising 10 high-fidelity (HF) simulations, carefully selected from a separate Latin hypercube that was not part of the training set.
Here, we will use {\mfbox} to denote the emulators using the GMGP model \citep{Ji:2021} with the graph structure in Figure~\ref{fig:nbody_plot}.
Section~\ref{subsec:results_dgmgp_accuracy} will show how {\mfbox}'s accuracy improves by adding an L2 node in $100 \Mpch$.
Section~\ref{subsec:results_boxsize} will show how {\mfbox}'s accuracy changed as a function of L2 box size, from $100 \Mpch$ to $256 \Mpch$.
Finally, Section~\ref{subsec:results_dgmgp_versus_nargp} show the runtime comparison between single-fidelity emulators, {\mfemu} (including AR1, NARGP) and {\mfbox}.



\subsection[MF-Box Accuracy (256 + 100 Mpc/h)]{{\mfbox} accuracy (256 + 100 $\Mpch$)}
\label{subsec:results_dgmgp_accuracy}

This section shows how the emulation error changed when a suite of small-box simulations is included as a second LF node, L2, through {\mfbox}.
More precisely, we use two LF nodes:
\begin{itemize}
   \item L1: $128^3$ simulations with $256 \Mpch$;
   \item L2: $128^3$ simulations with $100 \Mpch$.
\end{itemize}
The information about the training simulations is summarized in Table~\ref{table:simulations}.

\begin{figure}
   \includegraphics[width=\columnwidth]{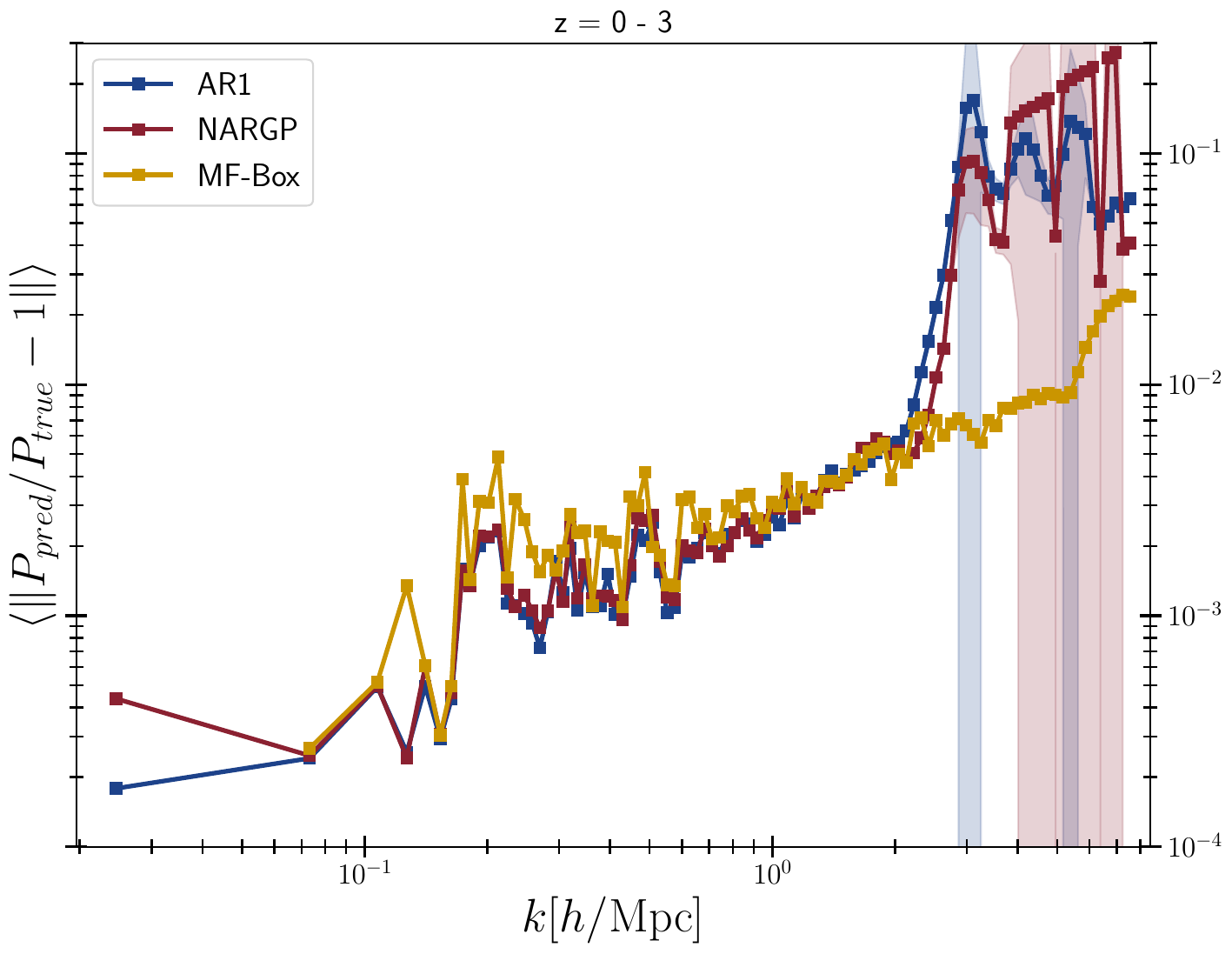}
   \caption{Relative errors averaged over $z = [0, 0.2, 0.5, 1, 2, 3]$ for different multi-fidelity models, AR1 (\textcolor{skyline}{blue}), NARGP (\textcolor{flatirons}{red}), and {\mfbox} (\textcolor{sunshine}{yellow}).
   The {\mfbox} model uses 60 L1 ($256 \Mpch$), 60 L2 ($100 \Mpch$), and 3 H ($256 \Mpch$) simulations for training.
   Both AR1 and NARGP use 60 L1 and 3 HF for training.
   The shaded area is the variance among different test simulations.
   }
   \label{fig:dgmgp_all_z_versus_k}
\end{figure}

Figure~\ref{fig:dgmgp_all_z_versus_k} shows the emulation error averaged over redshift bins, $z \in [0, 3]$, by using different multi-fidelity models, AR1, NARGP, and {\mfbox}.
All three models perform similarly at large scales ($k < 2 \hMpc$).
The main difference is {\mfbox} performs better at $k \geq 2 \hMpc$ while AR1 and NARGP have an error bump at $10\%$ level.

\begin{figure*}
   \includegraphics[width=2\columnwidth]{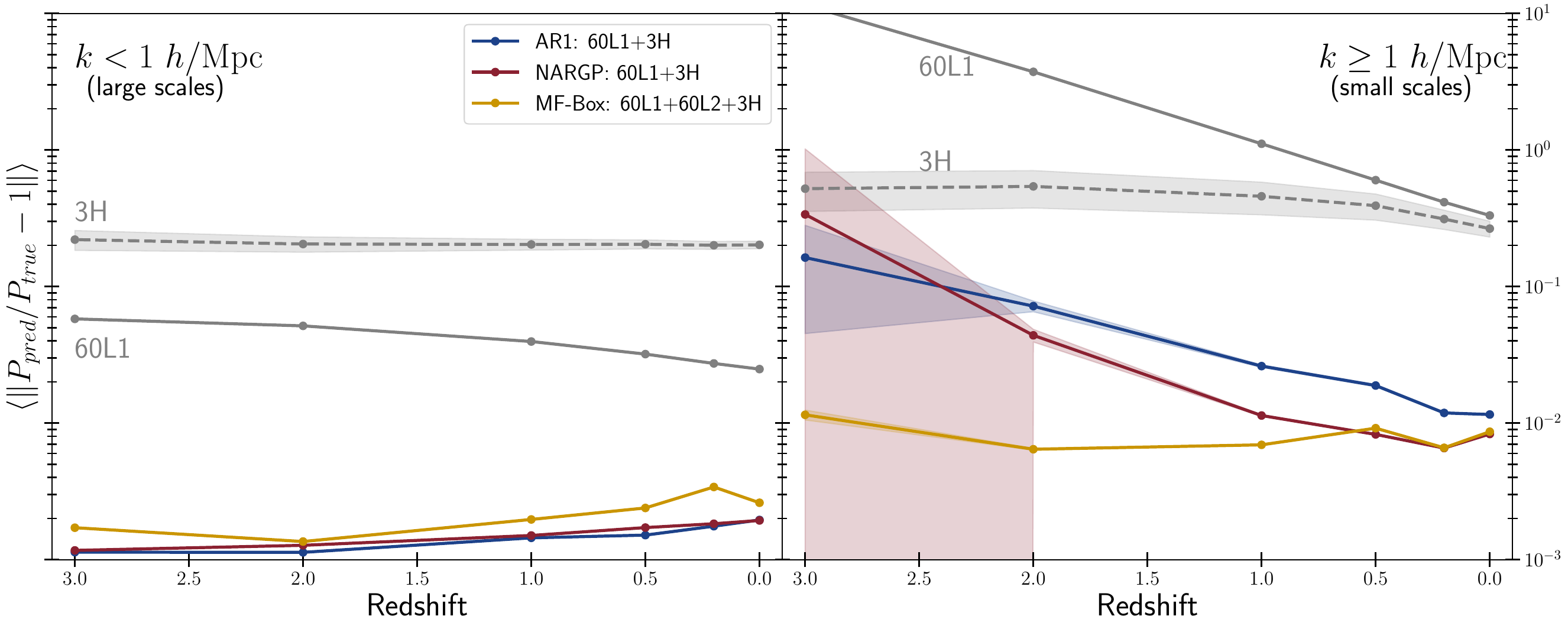}
   \caption{Relative errors averaged over all k modes (split into large and small scales) for different multi-fidelity models (AR1 (\textcolor{skyline}{blue}), NARGP (\textcolor{flatirons}{red}), and {\mfbox} (\textcolor{sunshine}{yellow})), broken down into different redshift bins.
   The grey dashed line is the HF-only emulator using 3 H simulations, and the solid grey line is the LF-only emulator using 60 L1 simulations.
   The shaded area is the variance among different test simulations.
   {\mfbox} improves the emulation at small scales at higher redshifts ($z \geq 1$).
   We do not include the variance of LFEmu (60L1) because the variance is too large.
   }
   \label{fig:dgmgp_all_k_versus_z_and_scales}
\end{figure*}

In the right panel of Figure~\ref{fig:dgmgp_all_k_versus_z_and_scales},
we can easily see the $10 \%$ error bump exists at $z = 1 - 3$ at small scales ($k \geq 1 \hMpc$).
The small-scale improvement in the right panel is not a surprise.
The additional low-fidelity node in a smaller box (L2) brings more accurate small-scale statistics than L1,
making {\mfbox} outperform AR1 and NARGP.
{\mfbox} stays $\simeq 1\%$ error within the redshift range $z \in [0, 3]$, in contrast to AR1 and NARGP where the error increases from $\simeq 1\%$ to $\simeq 20\%$ (from $z = 0$ to $z = 3$).

The bump in interpolation error in AR1 and NARGP at $z > 1$ is due to the feature at the initial inter-particle spacing at these redshifts, corresponding to the initial particle grid, as mentioned in \cite{Ho:2022}. The mean particle spacing of the initial condition appears as a delta function in the matter power spectrum at high redshift.
This feature eventually disappears, erased by gravitational interactions.
The L2 and high fidelity box, however, both have a smaller mean inter-particle spacing and thus show the delta function on smaller scales, beyond those we wish to emulate. Using the information the L2 simulations provide, {\mfbox} is able to maintain similar accuracy across $z \in [0, 3]$.

The left panel of Figure~\ref{fig:dgmgp_all_k_versus_z_and_scales} shows the redshift trend at large scales,
indicating no significant difference between AR1, NARGP, and {\mfbox}.
The slightly worse accuracy in {\mfbox} is probably because {\mfbox} has more hyperparameters to fit, making it slightly more difficult to reach $\sim 0.1 \%$ accuracy.


\begin{figure*}
   \includegraphics[width=2\columnwidth]{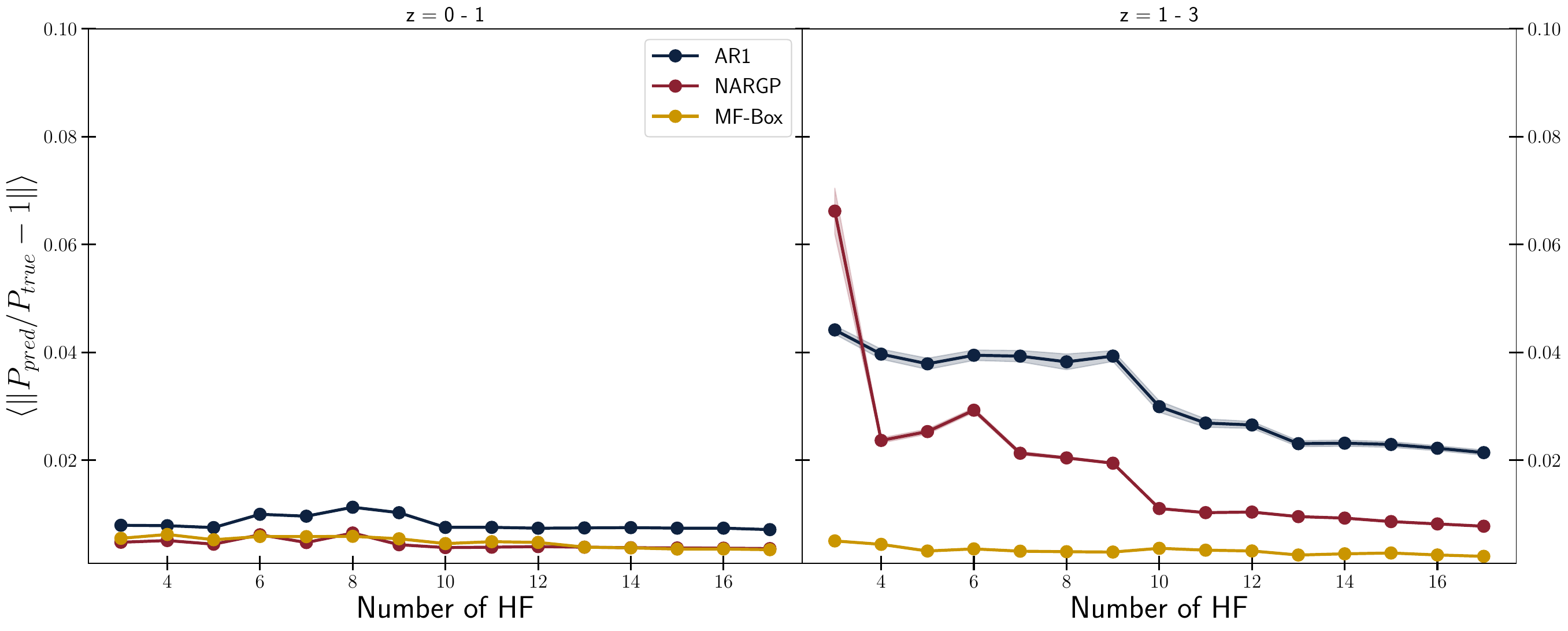}
   \caption{
   Relative error as a function of the number of HF training points for different multi-fidelity methods: AR1 (\textcolor{skyline}{blue}), NARGP (\textcolor{flatirons}{red}), and {\mfbox} (\textcolor{sunshine}{yellow}).
   The range of the number of HF points is relatively small, so the error estimate trend is unclear.
   However, in general, the emulation error decreases with more HF points.
   (Left) Averaged relative error for $z \in [0, 0.2, 0.5]$.
   (Right) Averaged relative error for $z \in [1, 2, 3]$.
     }
   \label{fig:vary_hf}
\end{figure*}

Figure~\ref{fig:vary_hf} shows the AR1, NARGP, and {\mfbox} accuracies as a function of the number of HF points, splitting into two redshift bins.
The left panel shows the accuracy averaged over the low redshift bins, $z \in [0, 0.2, 0.5]$, where NARGP and {\mfbox} perform similarly and outperform the AR1 model.
It is not a surprise that NARGP and {\mfbox} perform similarly since {\mfbox} is an extension of NARGP.

The left panel of Figure~\ref{fig:vary_hf} shows that the error is almost flat as a function of HF points. In Section~\ref{sec:budget}, we showed that the emulator error is a power-law function of the number of training points.
Here, the emulation accuracy is likely limited by the intrinsic accuracy of our $ 512^3 $ HF simulations, so it is hard to get improvement at the sub-percent level.\footnote{As discussed in \cite{Ho:2022}, our HF power spectra are $\sim 0.1 - 10\%$ error compared with EuclidEmulator2.}
The right panel of Figure~\ref{fig:vary_hf} shows that {\mfbox} performs better than the other two models by a factor of $\sim 5 - 10$.

\begin{figure*}
   \includegraphics[width=2\columnwidth]{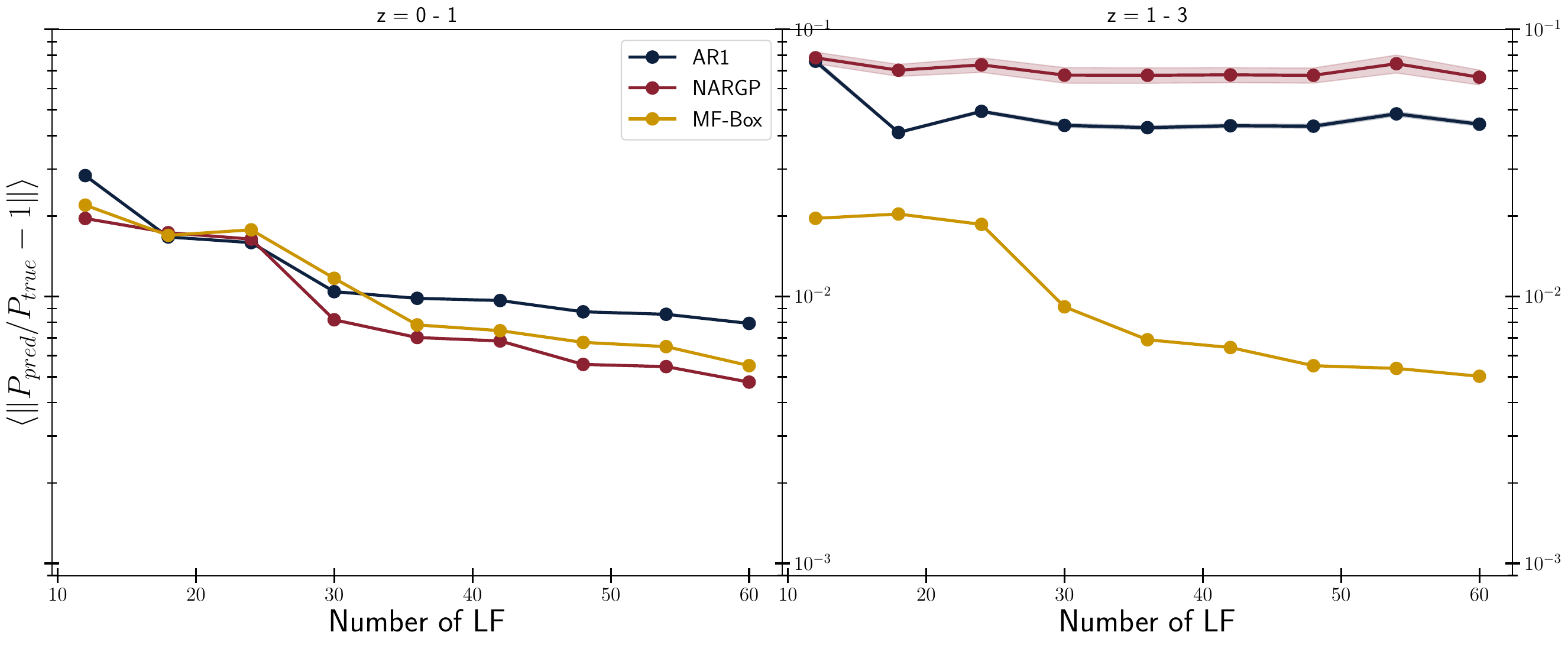}
   \caption
   {
   Relative errors for AR1 (\textcolor{skyline}{blue}), NARGP (\textcolor{flatirons}{red}), and {\mfbox} (\textcolor{sunshine}{yellow}) as a function of LF points, splitting into two redshift bins.
   (Left) Averaged error for $z \in [0, 0.2, 0.5]$.
   (Right) Averaged error for $z \in [1, 2, 3]$.
   }
   \label{fig:vary_lf}
\end{figure*}

Figure~\ref{fig:vary_lf} shows the averaged emulation error as a function of LF points.
We see a mild improvement at low-redshift bins (left panel) by adding more LF points for all three models.
At the higher redshift bins (right panel),
AR1 and NARGP cannot be easily improved by adding more LF training simulations.
This is likely because the error is dominated by the delta function in L1 at small scales.
{\mfbox} achieves an average error at the $1\%$ level with 30L1+30L2+3HF, as expected from Section~\ref{sec:budget}.

In summary, we show that the improvement of {\mfbox} happens at small scales ($k > 2 \hMpc$) at the higher redshift bins ($z \in [1, 2, 3]$).
This is primarily because the L1 node at these redshifts has the delta function feature from the initial particle grid dominating on small scales.


\subsection{Emulation with various box sizes}
\label{subsec:results_boxsize}

In Section~\ref{subsec:results_dgmgp_accuracy}, we have learned that we can achieve better emulation performance by incorporating a low-fidelity node in a smaller box.
This section examines how {\mfbox}'s emulation error changed as a function of the L2 box size.

\begin{figure*}
   \includegraphics[width=2\columnwidth]{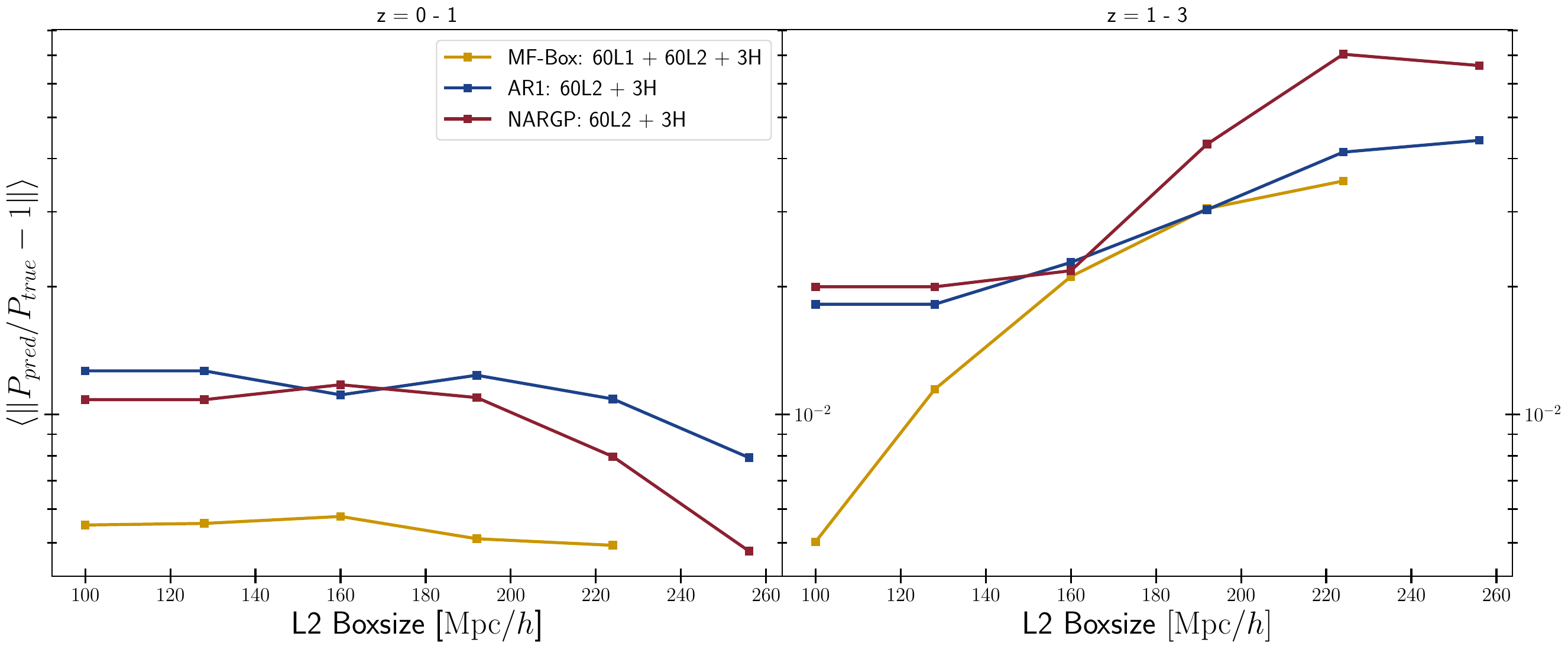}
   \caption{Relative errors of multi-fidelity emulation as a function of L2 boxsize, for AR1 (\textcolor{skyline}{blue}), NARGP (\textcolor{flatirons}{red}), and {\mfbox} (\textcolor{sunshine}{yellow}).
   Note that we use L2 instead of L1 for AR1 and NARGP models.
   }
   \label{fig:boxsize_split_z}
\end{figure*}

Figure~\ref{fig:boxsize_split_z} shows the emulation error as a function of L2 box size, averaging over all $k$ bins and splitting into two redshift bins.
We include AR1, NARGP, and {\mfbox}.
In this section, we use the L2 node as the LF node for both AR1 and NARGP.
The left panel shows the error at the low-redshift bin ($z \in [0, 0.2, 0.5]$).
AR1 and NARGP have $< 1 \%$ error with L2 $= 256 \Mpch$, but the error gets worse when the L2 box size becomes smaller due to the cosmic variance at large scales.
On the other hand, {\mfbox} error stays flat for $\mathrm{L2} \in [100, 224] \Mpch$.

The right panel of Figure~\ref{fig:boxsize_split_z} shows the error versus L2 box size at the high-redshift bin, $z \in [1, 3]$.
All models show a decrease in error using a smaller L2 box size in training.
This is mainly due to the feature at the initial inter-particle spacing mentioned in Section~\ref{subsec:results_dgmgp_accuracy}.
If a smaller L2 is used, the feature moves to smaller scales, away from those we are emulating, causing a decline of error from the large L2 box to the small L2 box size.

\begin{figure}
   \includegraphics[width=\columnwidth]{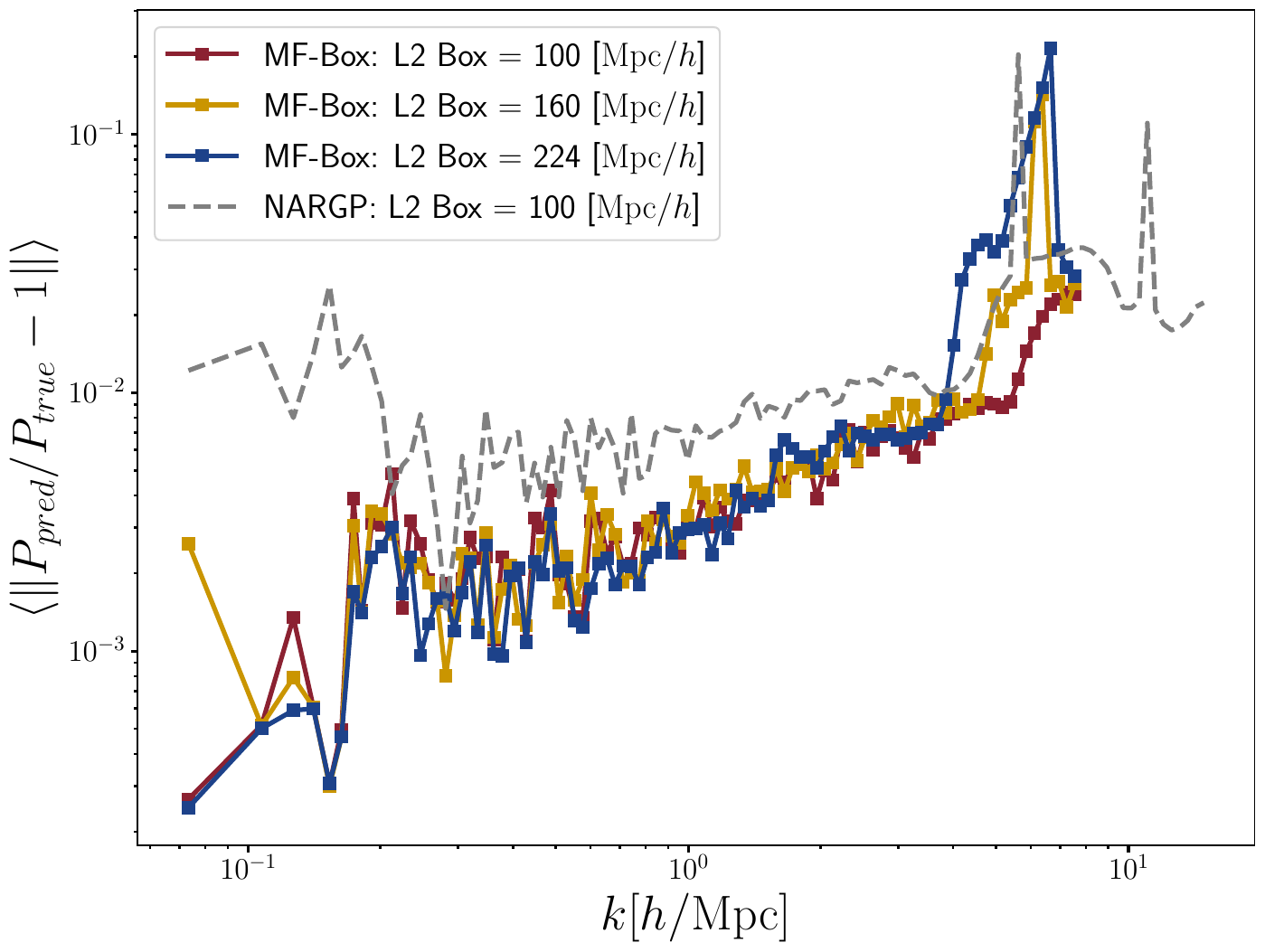}
   \caption{Relative errors averaged over redshift bins, as a function of k modes.
   {\mfbox} with $224 \Mpch$ L2 (\textcolor{skyline}{blue}),
   {\mfbox} with $160 \Mpch$ L2 (\textcolor{sunshine}{yellow}),
   and {\mfbox} with $100 \Mpch$ L2 (\textcolor{flatirons}{red}).
   The gray dashed line is the NARGP model uses $100 \Mpch$ L2.
   }
   \label{fig:boxsize_dgmgp_all_z_versus_k}
\end{figure}

To help visualize the performance change on different scales, we show in Figure~\ref{fig:boxsize_dgmgp_all_z_versus_k} the emulation error as a function of $k$, averaged over all redshift bins.
As Figure~\ref{fig:boxsize_dgmgp_all_z_versus_k} shows, for different L2 sizes, {\mfbox} accuracy only changes at the small scales with $k > 3 \hMpc$.
This is not a surprise because all {\mfbox} models share the same L1 node ($128^2$ simulations in $256 \Mpch$), and thus the emulation at large scales stays the same.
The NARGP shown in Figure~\ref{fig:boxsize_dgmgp_all_z_versus_k} uses L2 with $100 \Mpch$ as a low-fidelity node.
Its performance is worse than {\mfbox} with L2 $= 100 \Mpch$ at all $k$ bins.

To sum up, the error of {\mfbox} changed as a function of L2 box size: using a smaller L2 can result in better {\mfbox} accuracy.
The improvement caused by L2 is mostly at small scales ($k > 2 \hMpc$) at higher redshift bins ($z = 1, 2, 3$).

\subsection{Runtime comparison}
\label{subsec:results_dgmgp_versus_nargp}

\begin{figure}
   \includegraphics[width=\columnwidth]{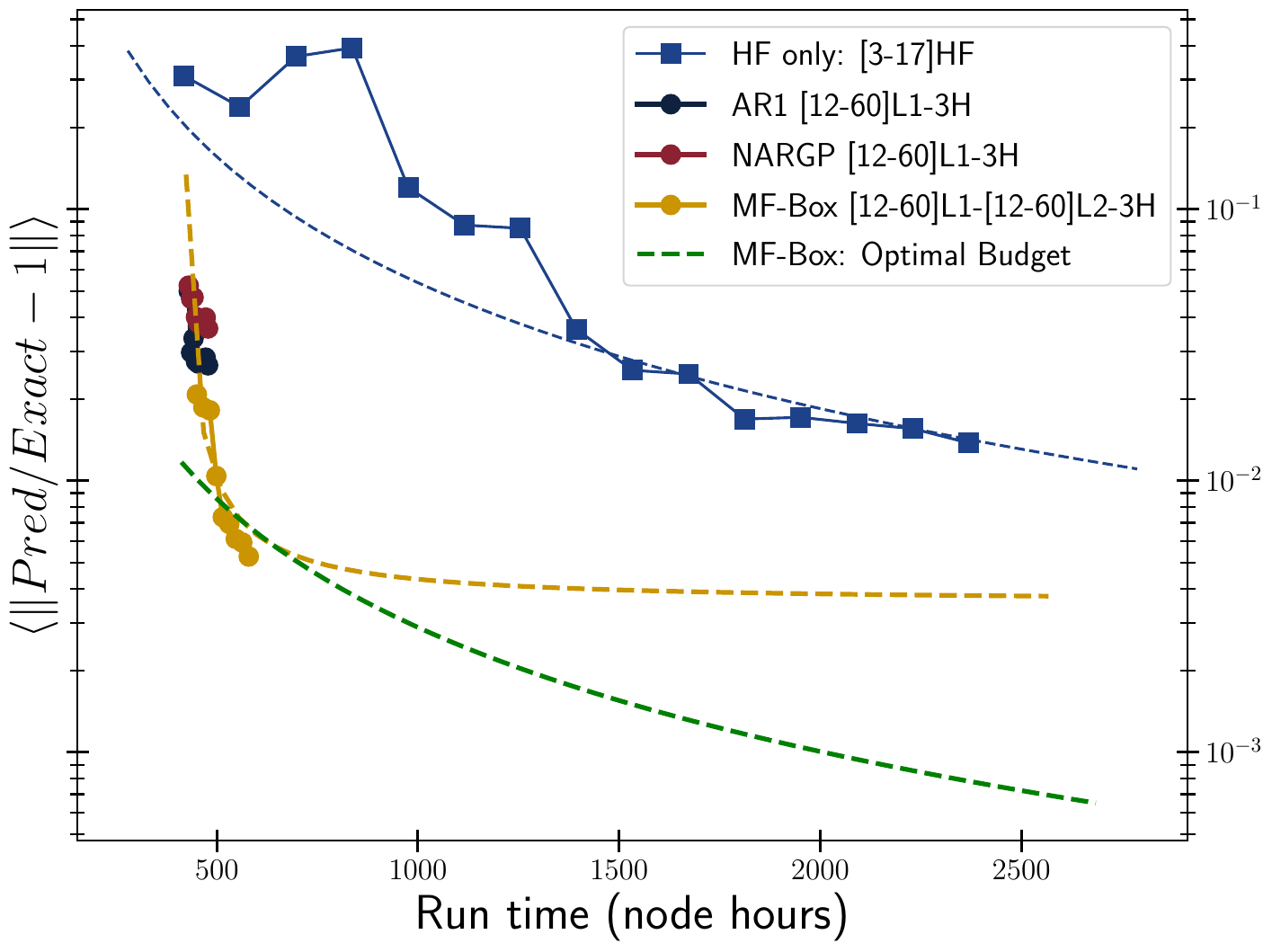}
   \caption{Runtime comparison in node hours.
   We average the error across redshift bins $z = [0, 0.2, 0.5, 1, 2, 3]$ and average across $k$ bins.
   AR1 and NARGP perform similarly to {\mfbox} at $z < 1$.
   Dashed lines are the predicted error based on the error function Eq~\ref{eq:error_function_approx}, which we inferred in Section~\ref{sec:budget}.
   }
   \label{fig:runtime}
\end{figure}

We will compare the costs of each method in this section.
Figure~\ref{fig:runtime} shows the error of different emulators as a function of node hours for the training simulations.
A similar compute time versus accuracy plot can be found in Figure 4 of \cite{Ho:2022}, albeit only for $z = 0$.
We performed the MP-Gadget simulations at High-Performance Computing Center (HPCC) at UC Riverside,\footnote{\url{https://hpcc.ucr.edu}}
each compute node has 32 intel Broadwell cores.

To understand Figure~\ref{fig:runtime}, we can start with the high-fidelity only emulators (\textcolor{midnight}{[3-11] HF}). This is the emulator we would train before we have multi-fidelity methods.
HF-only emulator shows a steady improvement with an increase in run time.
However, the error gradient gets flatter with more training points,
indicating the difficulty of improving an emulator at a highly accurate regime.

This trend is intuitive because the error of an emulator roughly scales as a power-law function, $\mathrm{(number \,of\, training \, points)}^{-\frac{\nu}{d}}$.
Each line in Figure~\ref{fig:runtime} is a segment of different power-law models.
In this view, we can see AR1 and NARGP follow two very similar trends, except one has a lower mean emulation error.


Switching the focus to {\mfbox}, we can see
the mean error of the power law is $\sim 6 - 8$ times better than AR1 and NARGP.
The error for both AR1 and NARGP plateaus, implying that adding new simulations will not increase the emulator's accuracy.
The only way to improve the emulation at a similarly good efficiency is using small-box simulations through {\mfbox}.

Recall the HF/L1 ratios in Figure~\ref{fig:powerspecs}. L1 is roughly at $\sim 5\%$ error at large scales.
On the other hand, the L2-only emulator is at $\sim 10\%$ error.
Using a {\mfbox},
the information carried by L1 and L2 is corrected to be at $\sim 0.5 \%$ level, which is a substantial improvement given that only 3 HF simulations are utilized to establish correlations between fidelities.

\section{Conclusions}
\label{sec:conclusions}

In this work, we show that our multi-fidelity emulation, {\mfbox} (model structure refers to Figure~\ref{fig:nbody_plot}, and simulation data refer to Table~\ref{table:simulations}), can combine simulations from different box sizes to achieve improved overall emulator accuracy.
{\mfbox} has a higher accuracy improvement per CPU hour than the multi-fidelity method with only one box size. The framework is adaptable to different simulation suites and emulation problems.

We summarize the key contributions of this work below:
\begin{enumerate}
    \item \textbf{Propose a new multi-fidelity emulation, {\mfbox}, combining information from different simulation box sizes}:
    Using the in-tree graph of GMGP \citep{Ji:2021}, we can fuse cheap low-fidelity simulations from multiple box sizes in one unified machine-learning model.
    Simulations in a large box capture large-scale statistics, while the simulations in a small box can improve small-scale statistics.
    Previously, the cheapest way to improve {\mfemu} was by increasing the particle load in the low-fidelity node, which scales as $\sim \mathcal{O}(\npart^3)$.
    {\mfbox} opens a new avenue to add additional information to the multi-fidelity emulation framework in a cheaper way.
    \item \textbf{Leverage accurate and systematic-free information from L2 to improve multi-fidelity emulation accuracy}:
    L2 provides unique information absent in L1, and also acts as a cross-check for L1.
    Systematic errors or unknown bugs in low-fidelity nodes can limit the effectiveness of multi-fidelity methods, as it relies on existing information.
    \cite{Ho:2022} identified such a limitation, noting that systematic errors present in the low-fidelity node can make achieving high accuracy difficult.
    {\mfbox} helps resolve the systematic in one low-fidelity node by introducing an additional L2 node without the systematic.
    It is worth noting that systematic errors may exist in both L1 and L2 nodes,
    but {\mfbox} can help mitigate these errors by cross-checking the information provided by two nodes, as long as the systematic errors are present at different scales.
    \item \textbf{Power-law analysis of emulation errors in multi-fidelity modeling with {\mfbox}:}
    In Section~\ref{sec:budget}, we present an error analysis of {\mfbox} models.
    We empirically estimate the emulation error function, which follows a power-law decay with respect to the number of training simulations. 
    This explains why it is difficult to improve single-fidelity emulators which are already percent-level accurate.
    Multi-fidelity emulation shows advantageous in reducing the overall cost and time required to achieve high accuracy.
    The estimated error function can also serve as a guide for optimizing resource allocation across fidelity nodes, facilitating the development of accurate emulators in a more efficient use of resources.
\end{enumerate}

{\mfbox} also opens up opportunities to experiment with different ways to implement multi-fidelity emulation in cosmology.
The second low-fidelity node, L2, can be anything that brings new information to a multi-fidelity emulator.
For example, it could be a node that runs using hydrodynamical simulations, or a node that uses a linear perturbation theory code.
One example could be L1 runs with dark-matter only simulations at high-resolution, L2 runs with hydrodynamical simulations at low resolution (and in a small box), and an HF node as hydrodynamical simulations at high-resolution.
This way, the cosmological dependence of the baryonic effects is captured by L2, and L1 gives us highly accurate gravitational clustering.
{\mfbox}, using a different box size in an additional low-fidelity node, is just a simple example to demonstrate the flexibility of this method.

The main remaining limitation of our multi-fidelity emulation framework is that the highest fidelity node must be in the training set, and encompass the largest box and highest resolution. In other words, our multi-fidelity framework cannot extrapolate to predict the results of a simulation with a resolution higher than the high-fidelity node.

Future applications of our multi-fidelity emulation include applying the {\mfbox} to the accurate high-resolution simulations, where the resolution can match the future experiments.
We may also apply {\mfbox} to different cosmological probes, especially applying to the beyond 2-point statistics, such as weak lensing peak counts and scattering transform coefficients.



\section*{Software}

We used the \texttt{GPy} \citep{gpy2014} package for Gaussian processes.
For multi-fidelity kernels, we moderately modified the multi-fidelity submodule from \texttt{emukit} \citep{Emukit:2019}.\footnote{\url{https://github.com/EmuKit/emukit}}
For the deep Graphical Model Gaussian Process (dGMGP) model, we used the code provided by \cite{Ji:2021}, which uses \texttt{GPy}.
For maxmin Sliced Latin Hypercube (SLHD), we use the R software \texttt{maximinSLHD} \citep{Ba:2015}.
We used the {\mpgadget} \citep{MPGADGET:2018} software for simulations.\footnote{\url{https://github.com/MP-Gadget/MP-Gadget}}
We generated customized dark matter-only simulations using Latin hypercubes through a modified version of \texttt{SimulationRunner}.\footnote{original: \url{https://github.com/sbird/SimulationRunner}; modified: \url{https://github.com/jibancat/SimulationRunnerDM}}.
Figure~\ref{fig:nbody_plot} is plotted using \texttt{gaepsi2}.\footnote{\url{https://github.com/rainwoodman/gaepsi2}}.
We also make use of the following python libraries: \texttt{matpltolib} \citep{Hunter:2007}, \texttt{numpy} \citep{Numpy:2020}, \texttt{scipy} \citep{Virtanen:2020}, and \texttt{pymc} \citep{Salvatier:2016}.

Our code is publicly available at \faicon{github-alt} \url{https://github.com/jibanCat/matter_emu_mfbox}, including an additional notebook example for the Tensorflow Probability\footnote{\url{https://www.tensorflow.org/probability}} \citep{TFP:2017} implementation of {\mfbox}.

\section*{Data Availability}

The simulation data are available at \faicon{github-alt} \url{https://github.com/jibanCat/MFBoxData}.
The emulator demo codes are available at \faicon{github-alt} \url{https://github.com/jibanCat/matter_emu_mfbox}.

\section*{Acknowledgements}

We would like to thank Dr. Simon Mak and Irene Ji for motivating this work and providing the Python code for the Graphical Model Gaussian Process (GMGP).
MFH thanks Yueying Ni for her guidance in creating visually appealing simulation figures for MP-Gadget using \texttt{gaepsi2}.
We thank Marjuka Ferdousi Lazin and Harini Venkatesan for their helpful discussions on combining different low-fidelity models in {\mfemu} during the early stages of the project.
We acknowledge the Cosmology from Home 2022 conference for providing a valuable virtual platform for discussions.
MFH thanks Sum (Mahdi) Qezlou, Yongda Zhu, Pak Kau Lim, Liz Finney, Reza Monadi, and Archana Aravindan for their valuable feedback on this work during the UCR's Physics and Astronomy Student Seminar (PASS).
MFH thanks Yanhui Yang for the valuable feedback.
MFH acknowledges the support of a National Aeronautics and Space Administration FINESST grant under No. ASTRO20-0022.
MAF is supported by a National Science Foundation Graduate Research Fellowship under grant No. DGE-1326120.
SB acknowledges funding from NASA ATP 80NSSC22K1897.
Computations were performed using the computer clusters and data storage resources of the HPCC, which were funded by grants from NSF (MRI-2215705, MRI-1429826) and NIH (1S10OD016290-01A1), and Frontera computing project at the Texas Advanced Computing Center (TACC), which was funded through National Science Foundation award OAC-1818253.

\bibliography{sample}



\label{lastpage}

\end{document}